%% file: main.tex
\documentclass[12pt,a4paper]{article} 

\usepackage[english]{babel}
\usepackage{amsfonts}
\usepackage{amsmath}
\usepackage{doi}
\usepackage{fullpage}
\usepackage{graphicx}
\usepackage{hyperref}
\usepackage{parskip} 	
\usepackage{setspace}
\usepackage{url}
\usepackage{float}
\usepackage{subcaption}
\usepackage{authblk}
\usepackage{fancyhdr}
 \usepackage{booktabs} 
\usepackage{natbib}
\setcitestyle{authoryear,open={(},close={)}} 

\title{Optimal Lattice Boltzmann Closures \\  through Multi-Agent Reinforcement Learning}
\author[1,2]{Paul Fischer}
\author[1]{Sebastian Kaltenbach}
\author[1]{Sergey Litvinov}
\author[3]{Sauro Succi}
\author[1,*]{Petros Koumoutsakos}
\affil[1]{Computational Science and Engineering Laboratory, Harvard University, Cambridge, MA02138, USA}
\affil[2]{ETH Zurich}
\affil[3]{Italian Institute of Technology, Rome}
\affil[*]{Corresponding author: petros@seas.harvard.edu}

\date{\today}

\begin{document}
\pagenumbering{gobble}
\maketitle
\pagestyle{fancy}

\begin{abstract}
The Lattice Boltzmann method (LBM) offers a powerful and versatile approach to simulating diverse hydrodynamic phenomena, spanning microfluidics to aerodynamics. The vast range of spatiotemporal scales inherent in these systems currently renders full resolution impractical, necessitating the development of effective closure models for under-resolved simulations. Under-resolved LBMs are unstable, and while there is a number of important efforts to stabilize them, they often face limitations in generalizing across scales and physical systems. We present a novel, data-driven, multiagent reinforcement learning (MARL) approach that drastically improves stability and accuracy of coarse-grained LBM simulations. The proposed method uses a convolutional neural network to dynamically control the local relaxation parameter for the LB across the simulation grid. The LB-MARL framework is showcased in turbulent Kolmogorov flows. We find that the MARL closures stabilize the simulations and recover the energy spectra of significantly more expensive fully resolved simulations while maintaining computational efficiency. The learned closure model can be transferred to flow scenarios unseen during training and has improved robustness and spectral accuracy compared to traditional LBM models. We believe that MARL closures open new frontiers for efficient and accurate simulations of a multitude of complex problems not accessible to present-day LB methods alone.
\end{abstract}

\newpage
\pagenumbering{arabic}
\setcounter{page}{2}

\section{Introduction}

Simulations of hydrodynamics are the cornerstone of key scientific and engineering endeavors of our times. Their applications range  from aerospace engineering \citep{farhat2000cfd} and automotive design \citep{aultman2022evaluation} to  climate modeling \citep{mirzaei2021cfd}, and astrophysics \citep{trac2003primer}. A significant component of these applications involve turbulent flows, which are inherently characterized by chaotic behavior and intricate interactions spanning a wide range of spatial and temporal scales. Resolving  all scales of these complex phenomena implies the use of direct numerical simulations (DNS) to capture the smallest relevant Kolmogorov scales \citep{Kolmogorov-theory} and the appropriate interaction between small and large scale vortices. However, the computational cost associated with DNS increases exponentially with the Reynolds number (Re), a dimensionless parameter that quantifies the intensity of turbulence in fluid flows \citep{Pope_2000}. For  engineering applications, such as airflow over aircraft with Reynolds numbers in the range of $10^7$ to $10^8$, DNS would demand computational resources on the order of $10^8$ to $10^{10}$ CPU hours, rendering direct simulations prohibitively expensive \citep{CFD-HPC-requirements}. Consequently, the quest to develop turbulence models that achieve a balance between accuracy and computational efficiency is one of the great challenges in the field of fluid dynamics \citep{NASA-vision-2030} and in particular its engineering applications.

In response to these challenges, turbulence modeling approaches have been developed, most prominently Large Eddy Simulations (LES)(\citep{Smagorinsky1963,Deardorff1970,leonard1975energy,rogallo1984numerical,kim2024early} and Reynolds-average Navier-Stokes (RANS) equations \citep{LAUNDER1974269,durbin2002perspective,duraisamy2019turbulence,moser2021statistical}. These models mitigate computational demands by avoiding the resolution of all pertinent scales and instead model the effects of the unresolved phenomena. LES resolves explicitly larger turbulent structures but  remains prohibitively expensive for several engineering applications. RANS models average turbulence effects and  have significantly lower computational demands but  often struggle to accurately predict critical flow phenomena essential for engineering design and analysis.\\
Recently, Lattice Boltzmann methods (LBM), derived from the kinetic theory of gases, have established themselves as computationally efficient, competitive solvers for a broad range of flow phenomena\citep{chen1998lattice,cercignani2002boltzmann,succi2018lattice,ratkai2019phase,LBM_Krueger,namburi2016crystallographic}. LBM provide a potent  alternatives to classical flow solvers based on the Navier-Stokes equations. Their inherent locality facilitates computational parallelization even in the case of complex geometries, while their exact conservation properties enable LBM to compete with higher-order spectral methods. The efficient computation of macroscopic moments improves the implementation of classical turbulence models. Moreover, LBM opens up possibilities for a new class of fully kinetic closure models, an area that remains largely unexplored \citep{chen2003extended,succi2018lattice}. More specifically, the resort to the kinetic formalism opens up the possibility of describing turbulent flows far from the statistical steady state, in which the very notion of eddy viscosity fails because of the lack of scale separation between the large and small eddies \cite{chen2004expanded}. At the same time, the accuracy of LBM depends on the appropriate form of its hydrodynamic closures\cite{ansumali2007hydrodynamics,biferale2017optimal,simonis2022temporal,freitas2025posteriori,hosseini2023entropic}.

Recent advances in machine learning (ML) have spurred a wealth of applications in fluid mechanics \citep{ ml-for-fm-review, karniadakis2021physics, gao2024generative}. Despite these successes ML methods often struggle to generalize effectively to out-of-distribution data\cite{koumoutsakos2024roads}. This limitation is particularly problematic in modeling physical behaviors where adherence to conservation laws, symmetries, and other fundamental principles is crucial. To address these challenges, hybrid models have emerged as a compelling solution by integrating classical numerical methods with machine learning techniques. This synergy leverages the strengths of both approaches while mitigating their respective weaknesses. Hybrid models have been successfully applied to areas such as fluid dynamics \citep{Brenner} and climate modeling \citep{Kochkov2024-GCM}. In the context of CFD, these hybrid approaches are often regarded as data-driven turbulence models. They offer the potential to develop more robust closures without relying heavily on heuristics and expert knowledge, which are typical limitations of classical turbulence models \citep{ML-for-LES-seminal, ml-for-LES-AndreaBeck, Brenner}. With regard to machine learning approaches for LBM, a residual network architecture has been used to learn forcing terms that improve the accuracy of LBM \citep{XLB}. Similarly, fully connected neural networks have been used to predict ghost relaxation rates for multi-relaxation-time LBM \citep{Lettuce} and to estimate bulk viscosity-linked relaxation rates \citep{bulk_neural_coll}. In addition, conservation laws and lattice symmetries have been explicitly embedded in neural networks to develop neural collision operators \citep{NLBM, towards-neurall-coll}. All of these approaches require a differentiable LBM solver and thus need an underlying LBM simulation that is stable and whose derivatives can be estimated.

Among the various ML techniques, reinforcement learning (RL) stands out as a distinct approach, often classified as a form of semi-supervised learning. RL's capability for self-learning has led to significant breakthroughs in fields such as control systems, robotics, and strategic games like Go and Chess \citep{Mnih2015-ControlAtari, Silver2016-AlphaGo, Silver2017-AlphaZero, Schrittwieser2020-MuZero}. This renders RL particularly advantageous for data-driven turbulence modeling \citep{RL-for-fluids-review}. In particular, RL does not require differentiable fluid solvers, can achieve long-term stability, and may eliminate the need for direct numerical simulation (DNS) data during the training process \citep{sanderse2024scientificmachinelearningclosure}. Early examples include RL-based strategies to optimize collective swimming behaviors \citep{collective-swimming1} and reduce drag \citep{collective-swimming2}. More recent studies have utilized RL to develop turbulence models by controlling dissipation coefficients within Smagorinsky subgrid-scale models. These efforts have successfully achieved accurate energy spectra for turbulent channel flows \citep{Bae2022-wall-bounded-flows} and isotropic turbulence \citep{Novati2021, KURZ2023-DRL-LES} without relying on DNS data. Additionally, RL has been employed to derive Reynolds stress models for wall-bounded flows \citep{Kim2022-wall-bounded-DRL}, and grid-based RL approaches have been developed to learn closure models that enhance coarse-grained simulations \citep{vonbassewitz2024closurediscoverycoarsegrainedpartial}.

In the present paper, we explore the fusion of reinforcement learning and Lattice Boltzmann methods to build novel hybrid Reinforced LBM (ReLBM) approaches. To the authors' best knowledge, no attempts have yet been made to enhance LBM with RL. We propose a multi-agent reinforcement learning algorithm that autonomously discovers control policies for the over-relaxation parameter. This algorithm successfully stabilizes under-resolved LBM simulations while accurately reproducing the energy spectrum of DNS simulations. Our model demonstrates robust generalization capabilities to extended roll-outs and higher Reynolds number simulations, as demonstrated for two-dimensional Kolmogorov flows and decaying turbulent flows. These results underscore the promise of ReLBM and lay the foundation for future advancements in this emerging field.

\section{Background}

\subsection{Lattice Boltzmann Method}

The foundation of LBM is the Lattice Boltzmann equation that  represents a fluid flow as a discrete set of particles moving on a regular grid:
\begin{equation}\label{eq:LBE}
    f_i(\boldsymbol{x} + \boldsymbol{c}_i\Delta t, t+\Delta t) - f_i(\boldsymbol{x},t) = \Omega_i(\boldsymbol{f}(\boldsymbol{x},t)).
\end{equation}
The equation describes the evolution of the discrete particle distributions $\boldsymbol{f} = (f_1,...f_b)^T$ corresponding to the discrete velocity $\boldsymbol{c}_i$ with $\Omega_i$ expressing the Boltzmann collision operator. Updates are performed by a collision step $f'_i = f_i + \Omega_i(\boldsymbol{f})$, and a streaming step $f_i(\boldsymbol{x} + \boldsymbol{c}_i\Delta t, t+\Delta t) = f_i(\boldsymbol{x},t) + f'_i$. The discrete velocities are chosen such that streamed populations exactly coincide with neighboring lattice sites. Various LBM formulations arise from different approximations of the collision operator. The most general is the Multi-Relaxation Time (MRT) method \citep{d2002multiple}, while the simplest, known as the Bhatnagar-Gross-Krook (BGK) approximation \citep{bhatnagar1954model}, leads to the LBGK equation:
\begin{equation}\label{eq:LBGK}
    f_i(\boldsymbol{x} + \boldsymbol{c}_i\Delta t, t+\Delta t) - f_i(\boldsymbol{x},t) = \omega \Delta t(f_i^{eq}-f_i)
\end{equation}
This approximation shifts from collision modeling to a linear relaxation at rate $\omega$ towards a local equilibrium distribution $f_i^{eq}$. A Chapman-Enskog analysis reveals that the LBGK equation approximates the Navier-Stokes equations with second-order accuracy, subject to a compressibility error of $\mathcal{O}(\mathrm{Ma}^2)$ \citep{chen1992recovery,benzi1992lattice}. The fluid’s dynamic viscosity depends on the relaxation rate as:
\begin{equation}\label{eq:viscosity-relaxation}
    \nu = \rho c_s^2\left( \frac{1}{\omega} - \frac{\Delta t}{2} \right),
\end{equation}
where $c_s$ denotes the speed of sound. Macroscopic observables (pressure, velocity, momentum flux tensor) are derived as moments of the distribution:
\begin{equation}\label{eq:moments}
        \rho = \sum_{i} f_i, \quad
        \rho u_a = \sum_{i} f_i c_{ia}, \quad
        \rho P_{ab} = \sum_{i} f_i c_{ia} c_{ib}.
\end{equation} 
Here $a,b$ indicate Cartesian coordinates in $d$-dimensions \citep{succi2018lattice, LBM_Krueger}.

\subsubsection{Closure Modeling}

Applying a reduction operator $\boldsymbol{\mathcal{A}}\boldsymbol{f} = \overline{\boldsymbol{f}}$ to the LBE \ref{eq:LBE}  introduces an unclosed term, $\overline{\Omega_i(\boldsymbol{f})}$. Closure modeling involves approximating this term parametrically:
\begin{equation}\label{eq:collision-closure}
    \Omega_{\theta, i}(\overline{\boldsymbol{f}}) \approx \overline{\Omega_i(\boldsymbol{f})}.
\end{equation}
Using the BGK approximation, the unclosed term is $\omega (\overline{f_i^{eq}(\boldsymbol{f})} - \overline{f_i})$, leading to potential closures for the equilibrium distribution and the relaxation rate:
\begin{equation}
    g_{\theta,i}^{eq}(\overline{\boldsymbol{f}}) \approx \overline{f_i^{eq}(\boldsymbol{f})},\quad
    \quad \omega_\theta(\overline{\boldsymbol{f}}) \approx \frac{\omega(\overline{f_i^{eq}(\boldsymbol{f})}-\overline{f_i})}{(f_i^{eq}(\overline{\boldsymbol{f}})-\overline{f_i})}.
\end{equation}
A key advantage of the lattice Boltzmann method (LBM) is its inherent satisfaction of physical conservation laws and symmetries. Consequently, it is advantageous to develop closure models that maintain these constraints. Although the approach of modeling the entire collision operator, as depicted in Eq. \ref{eq:collision-closure}, represents the most comprehensive method, it does not depend on the BGK approximation and is consequently subject to numerous constraints. Therefore, we have chosen to initially address the model with the fewest constraints $\omega_\theta(\overline{\boldsymbol{f}})$ \citep{succi2018lattice}.

\subsubsection{Entropic methods}

The Entropic Lattice Boltzmann Method (ELB) \cite{hosseini2023entropic} addresses instabilities, common in high Mach number and low viscosity flows, by enforcing a discrete H-theorem. Here, $\omega^* = \alpha \beta$ is redefined, with $\beta$ ensuring the viscosity remains accurate and $\alpha$ being analytically computed to satisfy the H-theorem. ELB has been shown to stabilize under-resolved simulations \citep{ELB-LES,ELB_vs_smagorinsky}, allowing it to function as a type of Large Eddy Simulation (LES) \citep{entropic-review}. The Karlin-Bösch-Chikatamarla (KBC) model extends ELB to the MRT model \citep{KBC}.

\subsection{Reinforcement Learning}

Reinforcement Learning (RL) is a form of approximate dynamic programming \citep{bertsekas2023course}. In RL, problems are generally formalized as a Markov Decision Process (MDP), defined by the tuple $\mathcal{M} = (\mathcal{S}, \mathcal{A}, \mathcal{P}, r, \mu, \gamma)$. In an MDP, an agent occupies states $s \in \mathcal{S}$, interacts through actions $a \in \mathcal{A}$, and transitions between states according to $\mathcal{P}$, where $p(s'|s, a)$ gives the probability of reaching state $s'$ from state $s$ by action $a$. For each action-state pair, the agent receives a reward $r(s, a)$. The initial state distribution is given by $\mu(s)$, and the discount factor $\gamma \in [0,1]$ balances immediate versus long-term rewards. RL seeks an optimal policy $\pi(a|s)$ that maximizes the expected discounted reward:
\begin{equation}
    \pi^* \in \mathop{argmax}_{\pi} \mathbb{E}_\tau \left[ \sum_{t=0}^T \gamma^t r(s_t,a_t) \right],
\end{equation}
where the expectation is over episodes $\tau \sim (\mathcal{M}, \pi)$, sampled as $\tau = (s_0, a_0, r_0, s_1, \ldots)$ \citep{Sutton-Barto}. Multi-agent RL (MARL) extends RL to partially observed stochastic games (POSGs), defined as $POSG = (\mathcal{I}, \mathcal{S}, {\mathcal{O}_i}, {\mathcal{A}_i}, \mathcal{P}, {r_i}, \mu, \gamma)$, with agents $i \in \mathcal{I}$ taking actions $a_i \in \mathcal{A}_i$, receiving rewards $r_i$, and observing states partially as $o_i \in \mathcal{O}_i \subseteq \mathcal{S}$. Any MDP can be translated into a POSG by assigning local rewards to agents, as in cooperative games where $r_i = r$ for all $i$ \citep{marl-book}.

\section{Methodology}

We propose a MARL framework to to stabilize under-resolved Lattice Boltzmann simulations inspired by the work on turbulence closure model discovery by \cite{Novati2021} as well as Grid-based Closure Modeling by \cite{vonbassewitz2024closurediscoverycoarsegrainedpartial}. Our hybrid model merges the locality and physical conservation properties of LBM with the function approximation capabilities of neural networks. Reinforcement learning’s posterior learning approach enables it to adaptively correct compounding errors over time, enhancing model stability and accuracy in complex systems \citep{sanderse2024scientificmachinelearningclosure}. Additionally, it can reproduce target statistics accurately without relying on costly DNS simulations during training, making it computationally efficient for large-scale problems. Flows were implemented with the XLB library \citep{XLB}, a Python-based LBM implementation using JAX \citep{jax2018github}. For reinforcement learning, we used the Tianshou library \citep{tianshou} and vectorized updates of independent, cooperative agents with fully convolutional networks as in \citep{vonbassewitz2024closurediscoverycoarsegrainedpartial}, allowing single-agent RL algorithms to update multiple agents efficiently.

\subsection{Kolmogorov Flow}\label{sec:kolmogorov-flow}

The two-dimensional Kolmogorov flow is a statistically stationary turbulent flow with periodic boundary conditions, driven by a sinusoidal forcing. It is often used as a benchmark example to test closure models. We replicated the flow configurations from \citep{Brenner} and converted them to lattice units (details in Appendix \ref{sec:kolmgorov-appendix}). Our LBGK simulation of Kolmogorov flow was implemented using the XLB library \citep{XLB}, with random initialization of the velocity field as described in \citep{initialization_kolmogorov}. After a burn-in period to reach statistical stability, we saved the resulting fields to initialize all future simulations. We consider a LBGK simulation on a Cartesian mesh of size $N = 2048^2$ as DNS, for a Kolmogorov flow at Reynolds number $Re = 10^4$. We aim to find a closure model for a coarse grid simulation (CGS) of size $N=128^2$ able to stabilize it and reproduce the target statistics of the DNS. For scaling the simulation parameters for varying the grid resolution at fixed Reynolds number we chose the convective scaling, to keep the velocity magnitude unchanged \cite{Brenner}.

\subsection{RL framework}

\begin{figure}[h]
    \centering
    \includegraphics[width=0.7\textwidth]{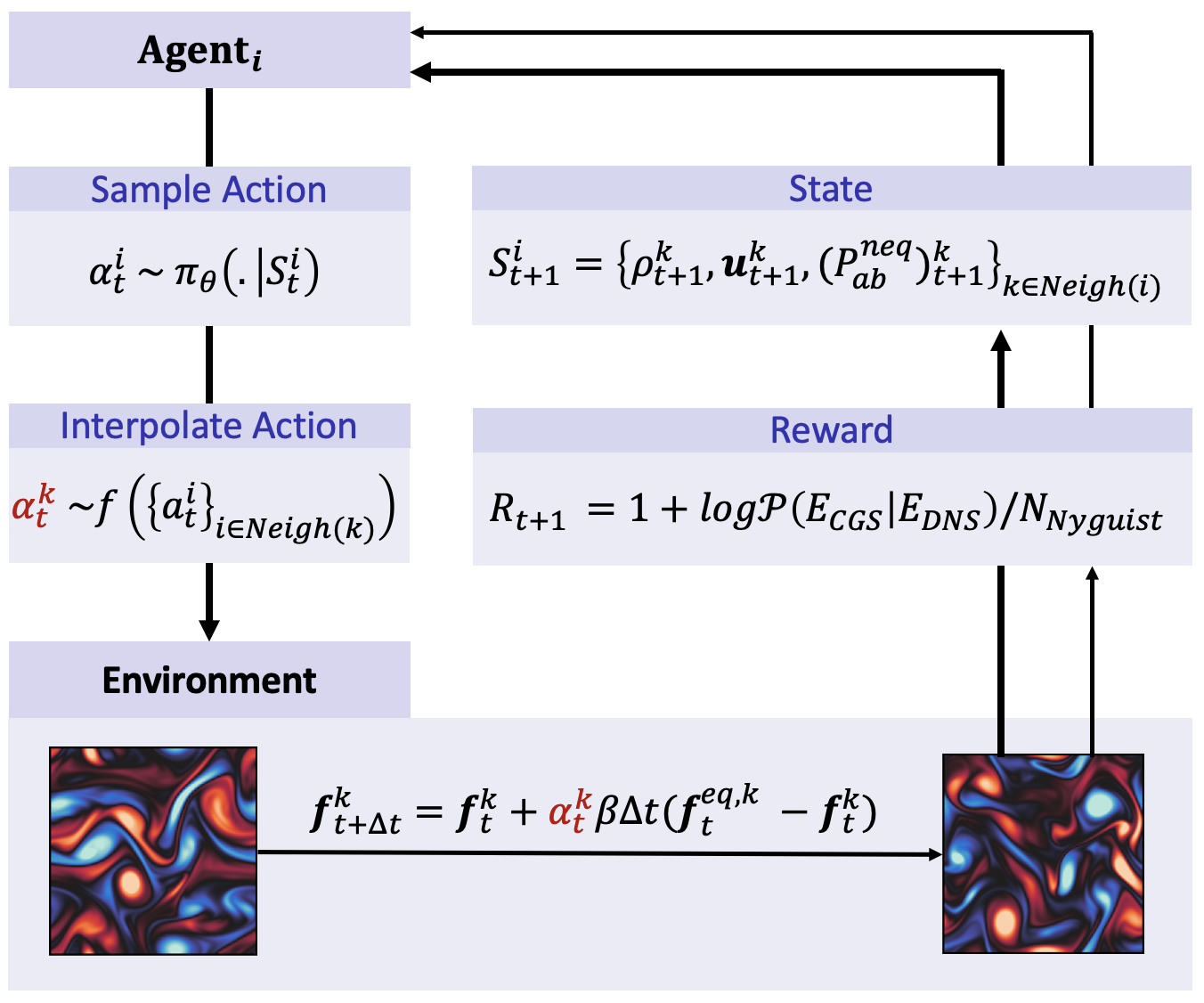}
    \hfill
    \caption{Reinforcement learning framework for adaptive control of the over-relaxation parameter in LBGK. Each agent samples an action from its policy, with actions interpolated across the grid and applied to the environment. Agents receive local state observations and a global reward based on the alignment of the coarse-grained simulation energy spectrum with the target DNS spectrum. }
    \label{img:rl-setup}
\end{figure}

The RL framework, illustrated in Figure \ref{img:rl-setup}, places $N_{agents}^2$ uniformly across the grid, with each agent having a perceptive field of size
\begin{equation}
    P = 
    \begin{cases}
        1 \quad &\mathrm{if} \quad N = N_{agents}\\
        N^2 \quad &\mathrm{if} \quad N_{agents}=1\\
        \left( \frac{N}{N_{agents}} + 1\right)^2 \quad &\mathrm{else}.
    \end{cases}
\end{equation}
Independent agents sample an over-relaxation parameter at their position $\boldsymbol{x}_i$ from a policy parametrized as a normal distribution $\alpha_i \sim \pi_{\theta_i}(\alpha|s_i) = \mathcal{N}(\mu_{\varphi_i}, \sigma_{\phi_i})$. The over-relaxation parameter at each grid point $k$ is then cubically interpolated from all agents actions. The resulting actions are applied to the environment, by performing $\Delta t_{RL}$ LBGK update steps with the local relaxation time $\alpha_k \beta$, inspired by the entropic Lattice Boltzmann method. The environment returns a state $S_i = \{ (\rho_k, \boldsymbol{u}_k, P_{ab}^{neq}) \}_{k \in Neigh(i)}$ to agent $i$, from a neighborhood defined by the agent's perceptive field. Closures are often computed as functions of the strain-rate tensor $S_{ab} = \frac{1}{2}\left(\frac{\partial u_a}{\partial x_b} + \frac{\partial u_b}{\partial x_a}\right)$ \citep{Pope_2000}. In LBM methods $S_{ab} = \frac{\omega}{\rho c_s^2}P_{ab}^{neq}$ \citep{succi2018lattice}, which is why we chose the second order non-equilibrium moment $P_{ab}^{neq}$ as part of the state. This setup was chosen to conserve the locality of LBM, allowing for easy parallelization even with the agent in the loop. All agents receive the same global reward 
\begin{equation}
    R = 1 + log\mathcal{P}(E_{CGS}|E_{DNS})/N_{Nyquist},
\end{equation}
 which is the log probability of sampling the energy spectrum of the coarse simulation from the desired energy spectrum distribution of the DNS. This is based on the observation from \citep{Novati2021}, that the log energy spectrum of the DNS is normally distributed. We can thus compute the mean and variance of the energy spectrum over time $(\mu_{DNS}, \Sigma_{DNS})$, from a few DNS simulations and do not need access to the expensive DNS during training at all. We note that the DNS energy spectra were computed on a down-sampled grid that matches the CGS grid resolution, thus $N_{Nyquist} = 64$. We scaled both spectra with $k^5$ such that the different wave numbers contribute equally to the loss, and scaled the resulting log spectra by a factor of 10. Hence $E' \leftarrow log(k^5 E)/10$, with which we can compute the probability:
\begin{equation}
    \mathcal{P}(E_{CGS}|E_{DNS}) \sim exp\left(-\frac{1}{2}(E'_{CGS}-\mu'_{DNS})^T (\Sigma'_{DNS})^{-1}(E'_{CGS}-\mu'_{DNS})\right).
\end{equation}

\begin{figure}[h]
    \centering
    \begin{subfigure}[t]{0.43\textwidth}
    \includegraphics[height=1.6in]{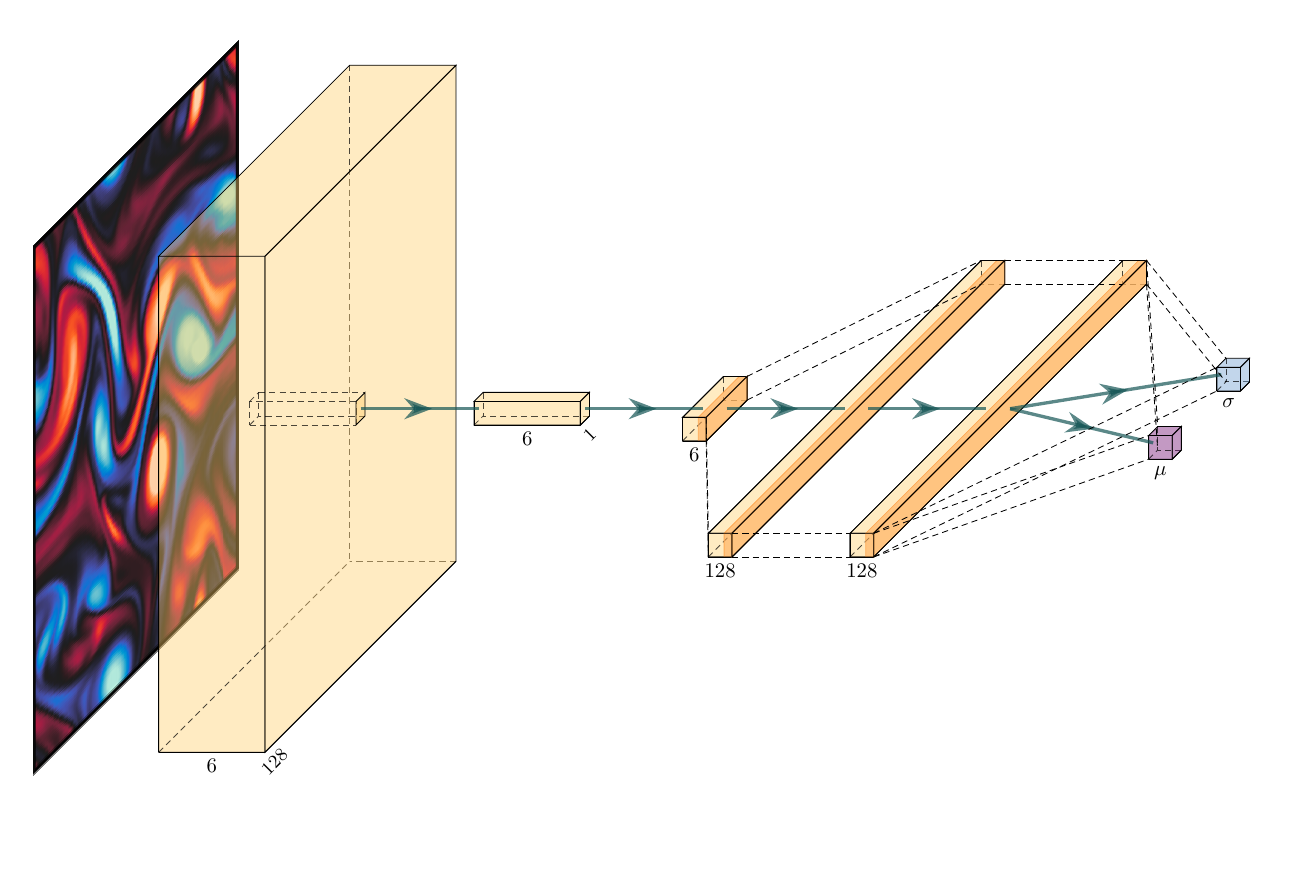}
        \caption{}
        \label{img:local-actor}
    \end{subfigure}
    \begin{subfigure}[t]{0.56\textwidth}
        \centering
        \includegraphics[height=1.6in]{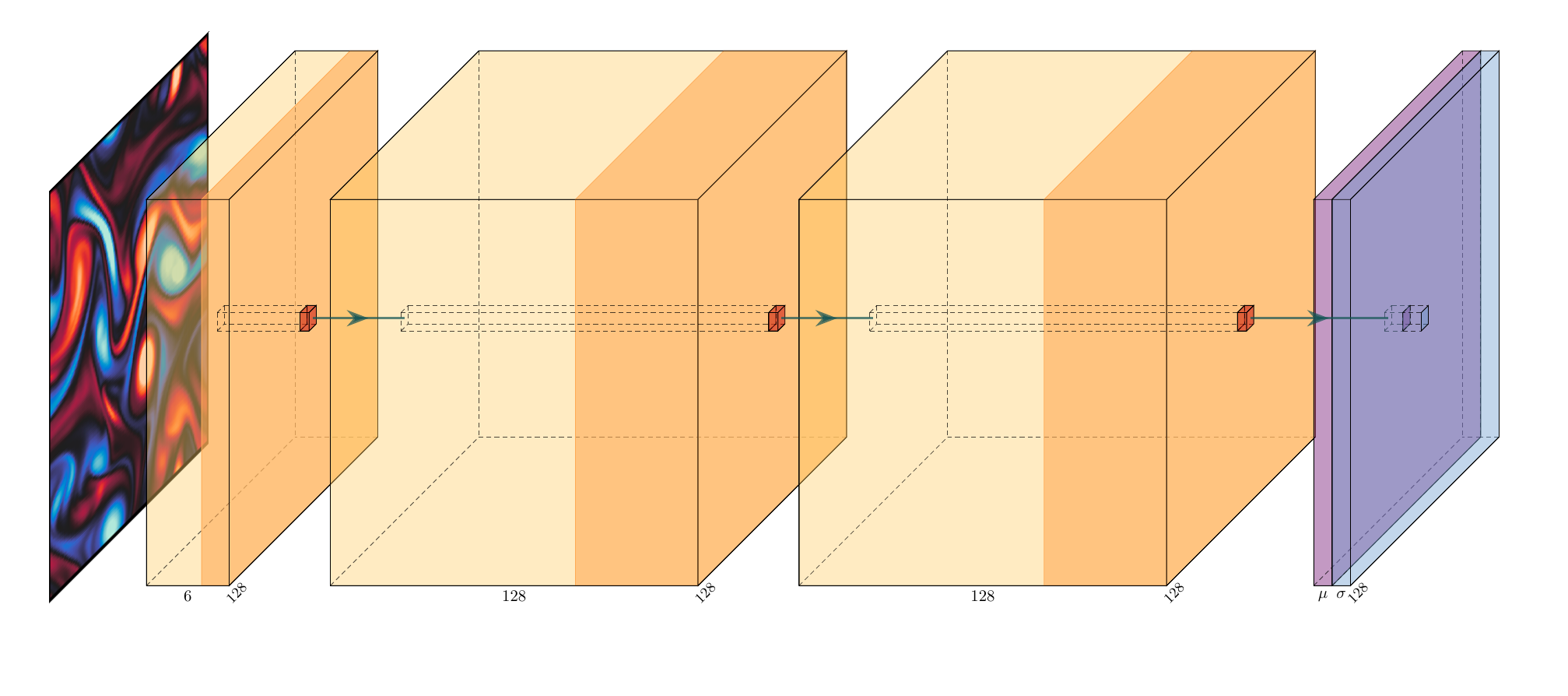}
        \caption{}
        \label{img:vectorized-actors}
    \end{subfigure}
    
    \caption{Example for a policy network parametrization in an environment with local and cooperating agents. Figure \ref{img:local-actor} shows the network architecture for a fully local agent. The agent only receives the state at its location which in this case is a six dimensional vector. The network then consists of two fully convolutional layers and has two output neurons, namely the mean $\mu$ and standard deviation $\sigma$. These parametrize a normal distribution from which the agent can sample actions $\pi_i(a|s)\sim \mathcal{N}(\mu, \sigma).$ Figure \ref{img:vectorized-actors} show the vectorized version of Fig. \ref{img:local-actor}. The experience of all agents is combined and handled as a single evaluation of a fully convolutional network.}
    \label{img:actor-network}
\end{figure}

\begin{figure}[h]
    \centering
    \includegraphics[width=0.9\textwidth]{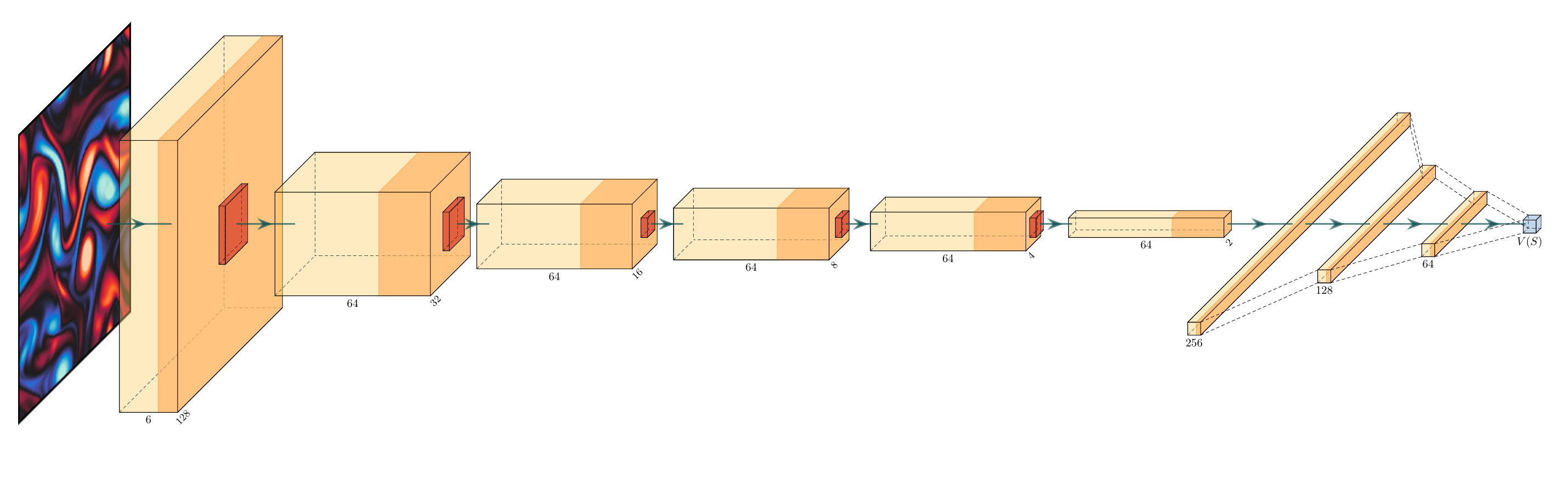}
    \hfill
    \caption{Neural network architecture for a centralized critic network used in the learning phase by the MARL PPO algorithm. The Network receives a state $S$ as input, which is compressed by six convolutional and three fully connected layers, and outputs a value function estimate $V(S)$.}
    \label{img:value-network}
\end{figure}

The agents are homogeneous and cooperate by sharing the reward $R_i = R$ while each is receiving local states. The globally optimal policy is obtained when all agents act according to the same locally optimal policy \citep{marl-book}. The agents share parameters $\pi_{\theta_i} = \pi_\theta \quad \forall i$, so the collective experience of all the agents during training is used to update the common policy. In \cite{vonbassewitz2024closurediscoverycoarsegrainedpartial} proposed an efficient way of computing this update by parametrizing the policy as a fully convolutional network (Fig. \ref{img:actor-network} shows an example architecture). This setup of fully decentralized agents (independent learning) has constant complexity with respect to the number of agents in contrast to centralized learning, which scales exponentially with the number of agents. However, one disadvantage of independent learning is that the presence of other agents is ignored and effectively treated as noise from the environment. This makes the environment nonstationary, often leading to convergence issues in training. To counteract this issue, we implemented a centralized training and decentralized execution method, using an actor-critic setup. The centralized critic receives the joint observation of all agents and can provide better value estimates in the training process (see Figure \ref{img:value-network}). This method combines the advantages of central and independent learning and still allows for a fully decentralized execution, since the critic is only used during training \citep{marl-book}. As an RL algorithm, we chose the multi-agent version of proximal policy optimization (PPO) \citep{PPO}, for its effectiveness in cooperative multi-agent settings reported in \citep{surprisingMAPPO} and its adaptability to continuous action spaces. We note that the fully convolutional parameterization of the policy networks allowed us to directly use the PPO implementation in Tianshou \citep{tianshou}.  

\subsubsection{Training setup}
We trained three different models with $N_{agents} \in \{1, 16, 128\}$ representing global (glob.), interpolating (interp.) and fully local (loc.) agents. The environment is terminated after a maximum of $10^4$ steps (i.e. non-dimensional T=113) but is stopped earlier if the simulations become unstable. This is the case if the realizability constraint $ 0 \leq f_i \leq 1 $ is violated or the velocity magnitude diverges. In this case, a truncation penalty $R_{truncated} = -100$ is added to the reward to penalize unstable behavior. We trained all agents for a maximum of $500$ epochs, where each epoch collects $1500$ state-action-reward tuples. We used the Adam optimizer \citep{ADAM} with an initial learning rate $\eta=10^{-3}$. A step factor of $\Delta t_{RL} = 8$ gave the best trade-off between model accuracy and noise for glob. and loc. agents. For the $16$ interpolating agents a step factor $\Delta t_{RL}=4$ was used, such that the receptive field contains all the surrounding states that are propagated to the agents location over the corse of one RL step. All hyperparameters are reported in \ref{sec:App_Hyp}.

\subsubsection{Testing setup}
We assessed the  models' stability, accuracy, and generalization, replicating the testing approach from \cite{Brenner}. All flows are initialized with seeds not seen during training. To assess long-term stability, we conducted simulations for double the training duration, totaling \(2 \times 10^4\) steps or a non-dimensional time of \(T = 227\). Throughout these simulations, we tracked the Pearson correlation coefficient between each model and the DNS, following the approach of \cite{Brenner}. Given the chaotic nature of turbulent flows, de-correlation over time is expected, even for two resolved simulations \citep{Pope_2000}. Consequently, this metric serves primarily as an indicator of stability, with unstable simulations showing an abrupt loss of correlation (see Figure \ref{img:corrs_1e4}). We further report the mean energy spectrum of models computed by averaging the spectra over the second half of the simulation (see Figure \ref{img:spectra_1e4}). For readability we did not include the standard deviations, however added them to the Appendix \ref{sec:std_spectra}. We then evaluated the performance on an unforced decaying turbulent flow, and lastly on a Kolmogorov flow at Reynolds number $Re=10^5$. To adjust the  models to local flow features at higher Reynolds numbers, we doubled the grid resolution for the $Re=10^5$ test case, as done by \cite{Brenner, Novati2021}. The fully convolutional networks allow for easy adaptations to different grids without any modifications. On all test cases we also report the performance of the KBC model and a BGK model with higher resolution.

\clearpage
\section{Experiments}

\begin{figure}[H]
    \centering
    \begin{subfigure}[t]{0.48\textwidth}
        \centering
    \includegraphics[height=1.8in]{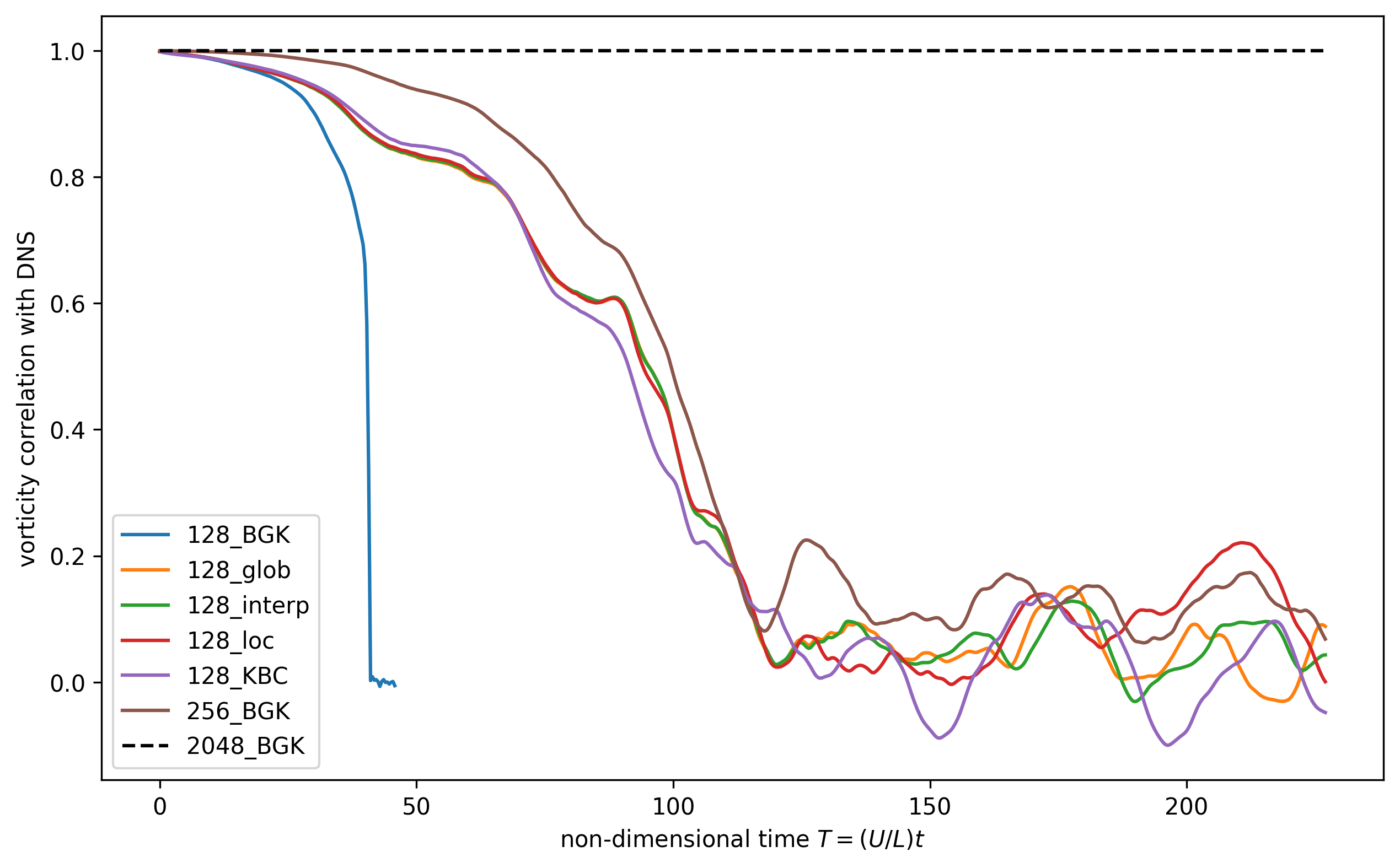}
        \caption{}
        \label{img:corrs_1e4}
    \end{subfigure}
    \begin{subfigure}[t]{0.48\textwidth}
        \centering
        \includegraphics[height=1.8in]{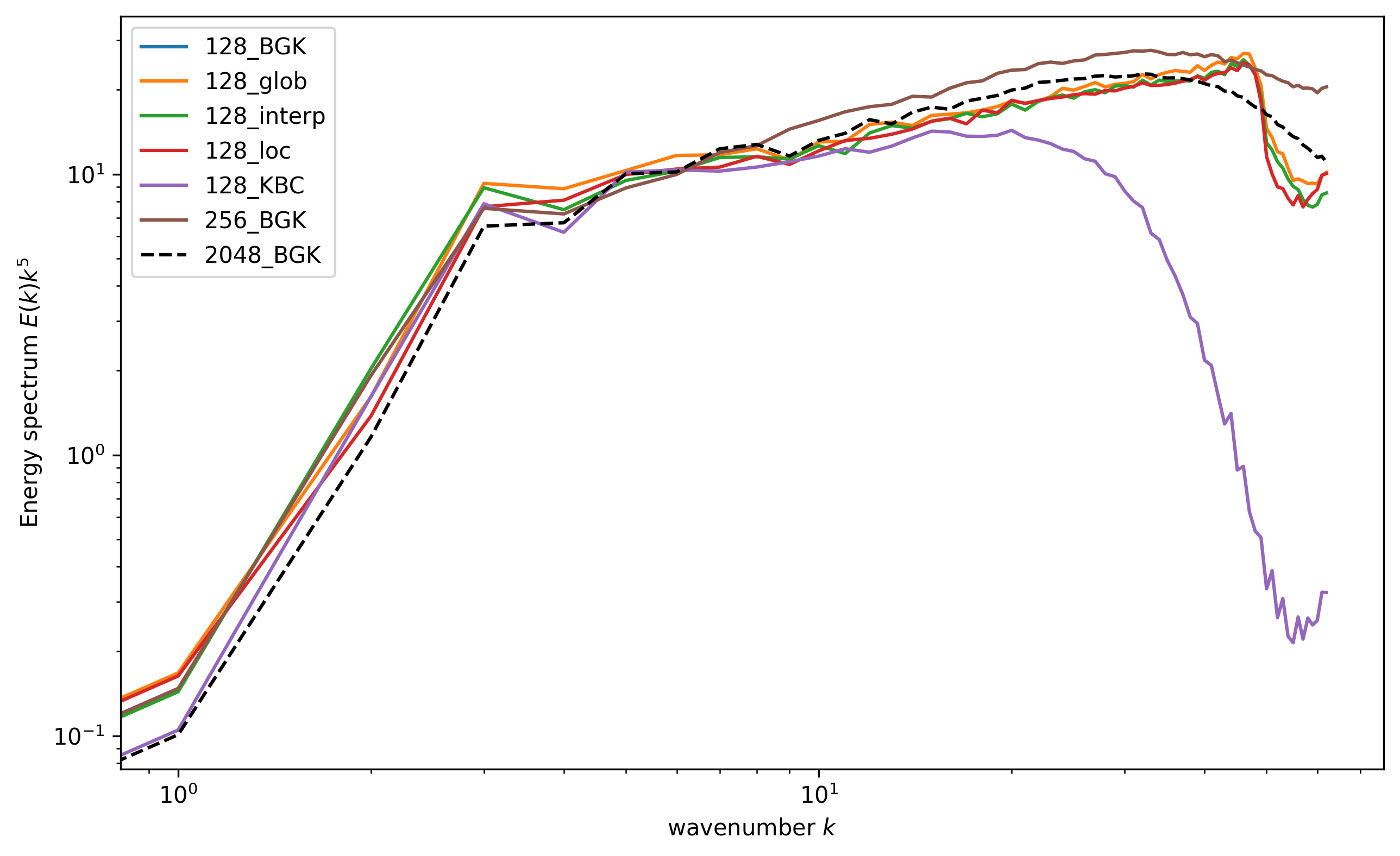}
        \caption{}
        \label{img:spectra_1e4}
    \end{subfigure}
    \caption{Evaluation of trained models on Kolmogorov flow at $Re=10^4$. (\ref{img:corrs_1e4}) show the vorticity correlation of models with the DNS. All three models are able to stabilize the simulation. (\ref{img:spectra_1e4}) shows the energy spectra scaled by $k^5$, averaged over the second half of the simulation $T\in[113, 227]$. All three models reproduce the target spectrum of the DNS, with small deviations at higher wave numbers.}
    \label{img:1e4_results}
\end{figure}

\begin{figure}[H]
    \centering
    \begin{subfigure}[t]{0.48\textwidth}
        \centering
    \includegraphics[height=1.8in]{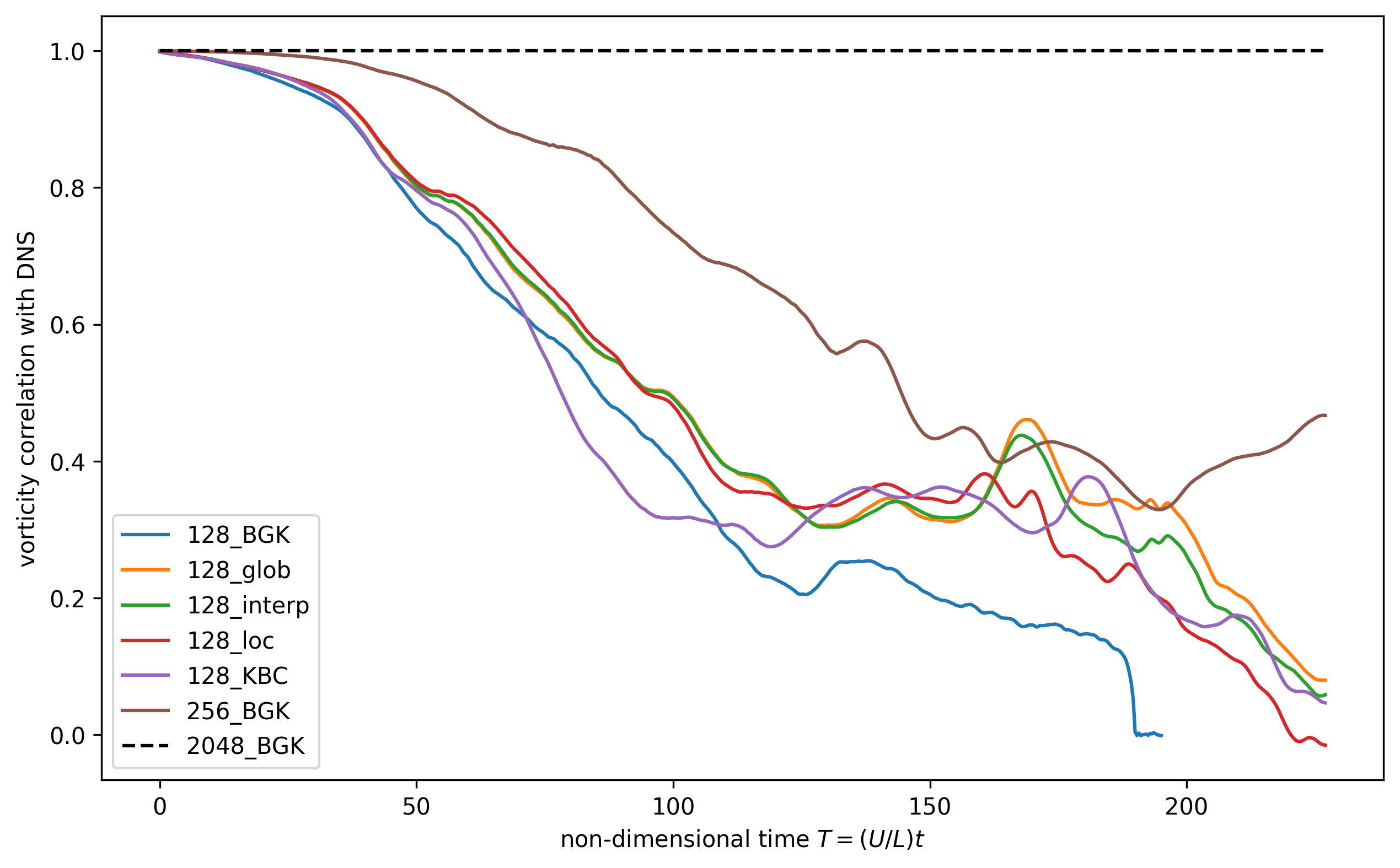}
        \caption{}
        \label{img:corrs_1e4_decay}
    \end{subfigure}
    \begin{subfigure}[t]{0.48\textwidth}
        \centering
        \includegraphics[height=1.8in]{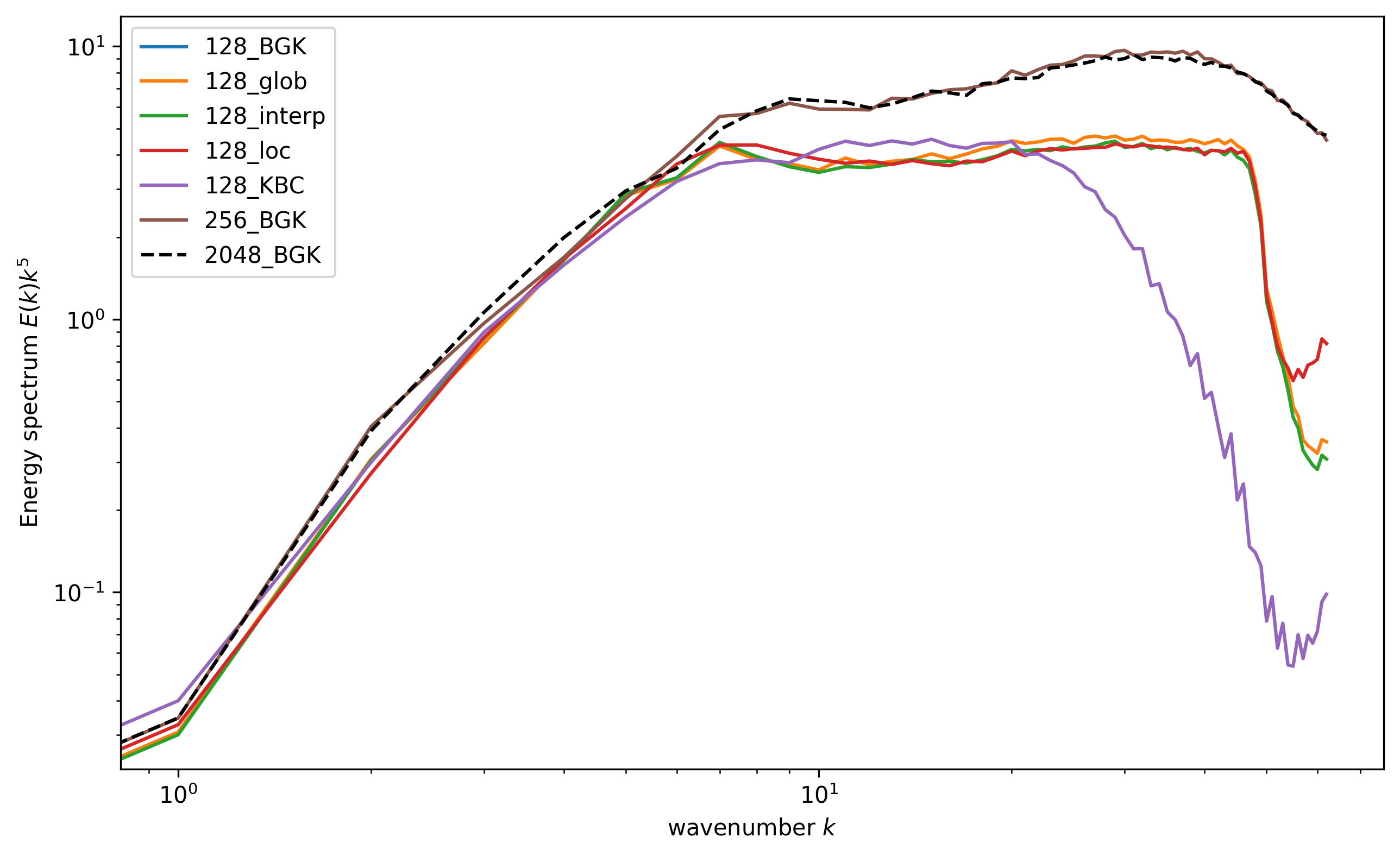}
        \caption{}
        \label{img:spectra_1e4_decay}
    \end{subfigure}
    \caption{Evaluation of trained models on an unforced decaying flow at $Re=10^4$.(\ref{img:corrs_1e4_decay}) show the vorticity correlation of models with the DNS. (\ref{img:spectra_1e4_decay}) shows the energy spectra scaled by $k^5$, averaged over the second half of the simulation $T\in[113, 227]$.}
    \label{img:1e4_decay_results}
\end{figure}

\begin{figure}[H]
    \centering
    \begin{subfigure}[t]{0.48\textwidth}
        \centering
    \includegraphics[height=1.8in]{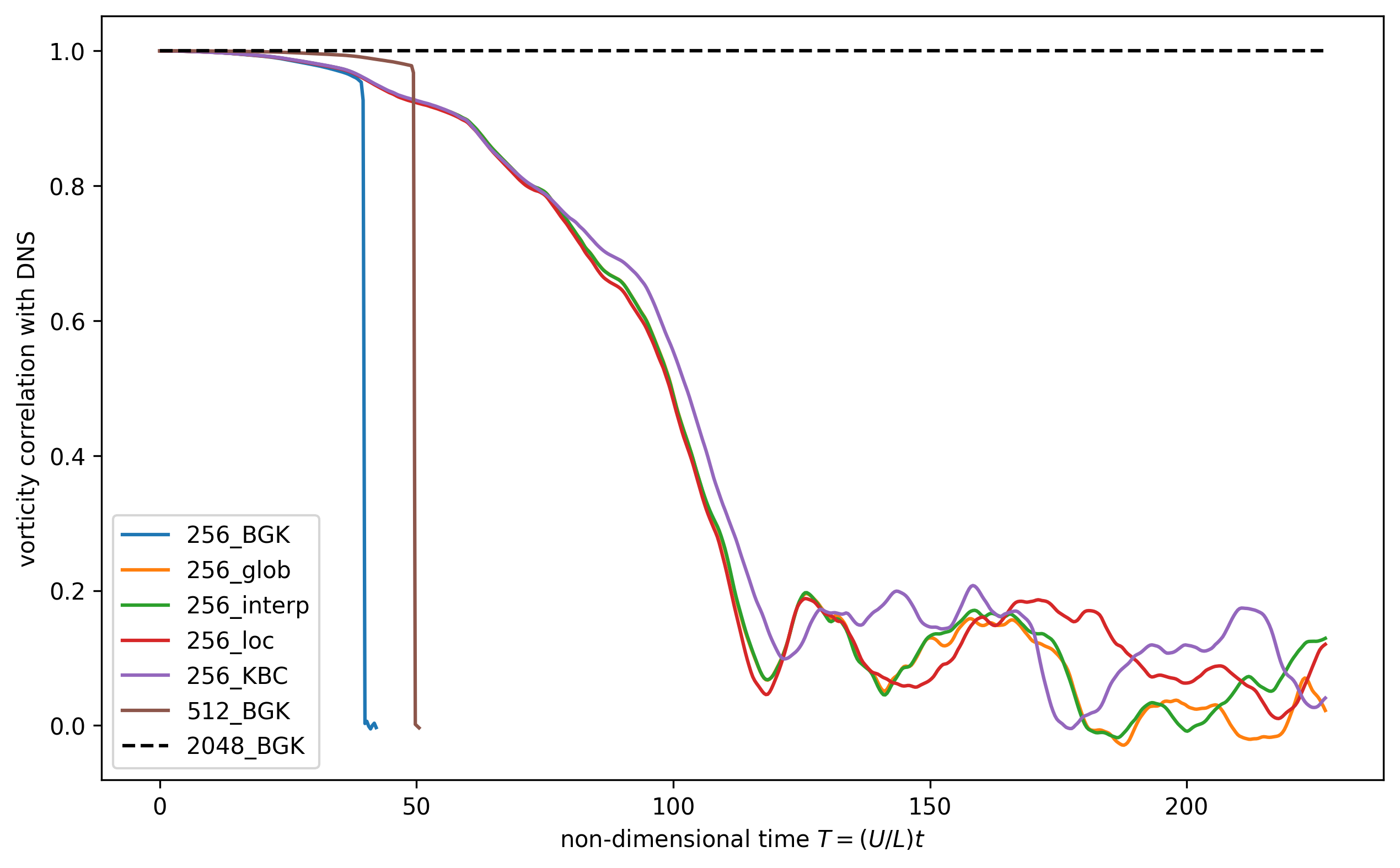}
        \caption{}
        \label{img:corrs_1e5}
    \end{subfigure}
    \begin{subfigure}[t]{0.48\textwidth}
        \centering
        \includegraphics[height=1.8in]{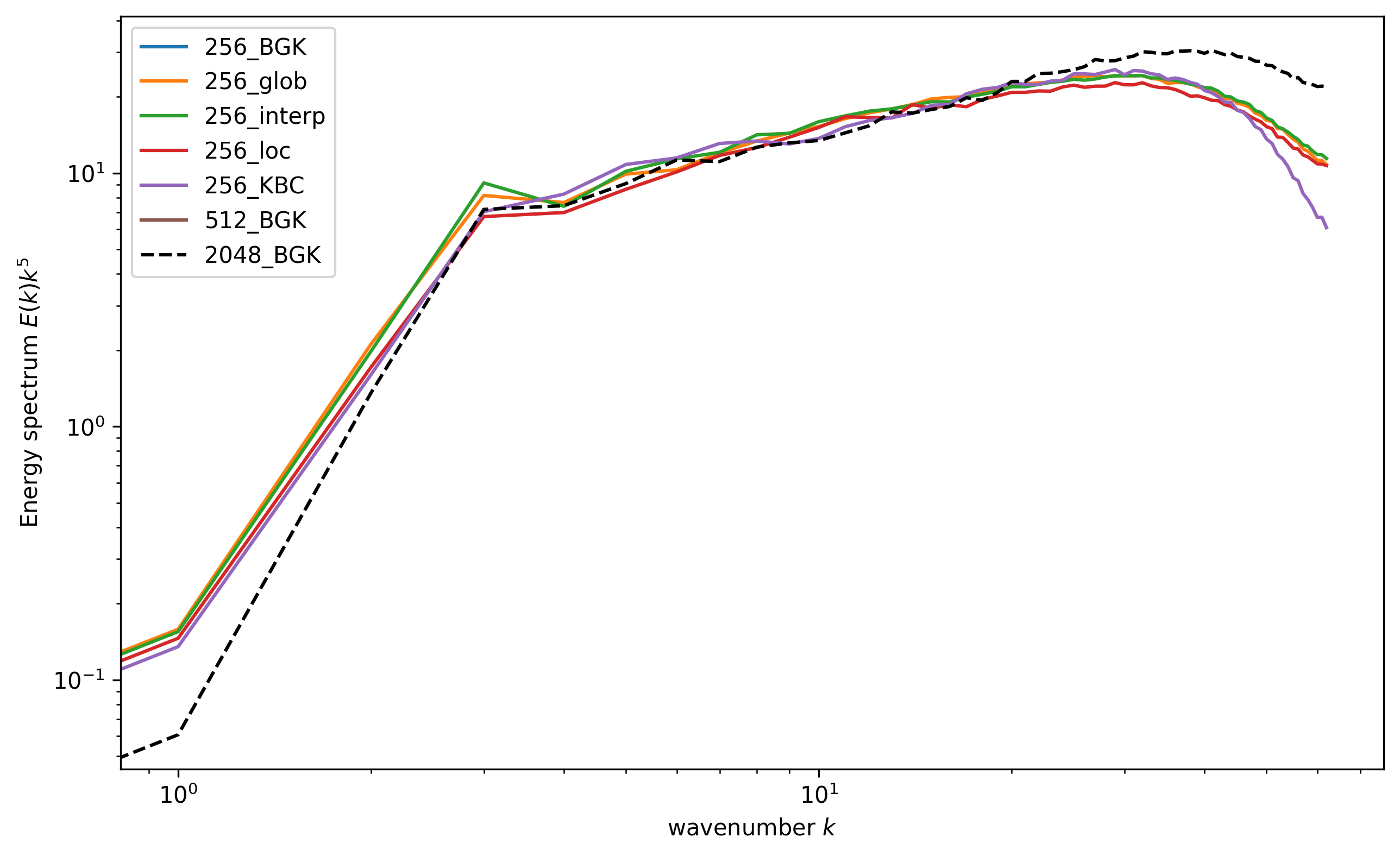}
        \caption{}
        \label{img:spectra_1e5}
    \end{subfigure}
    \caption{Evaluation of trained models on Kolmogorov flow at $Re=10^5$. (\ref{img:corrs_1e5}) show the vorticity correlation of models with the DNS. (\ref{img:spectra_1e5}) shows the energy spectra scaled by $k^5$, averaged over the second half of the simulation $T\in[113, 227]$.}
    \label{img:1e5_results}
\end{figure}

\begin{figure}[H]
    \centering
    \begin{subfigure}[t]{0.32\textwidth}
    \includegraphics[width=1.0\textwidth]{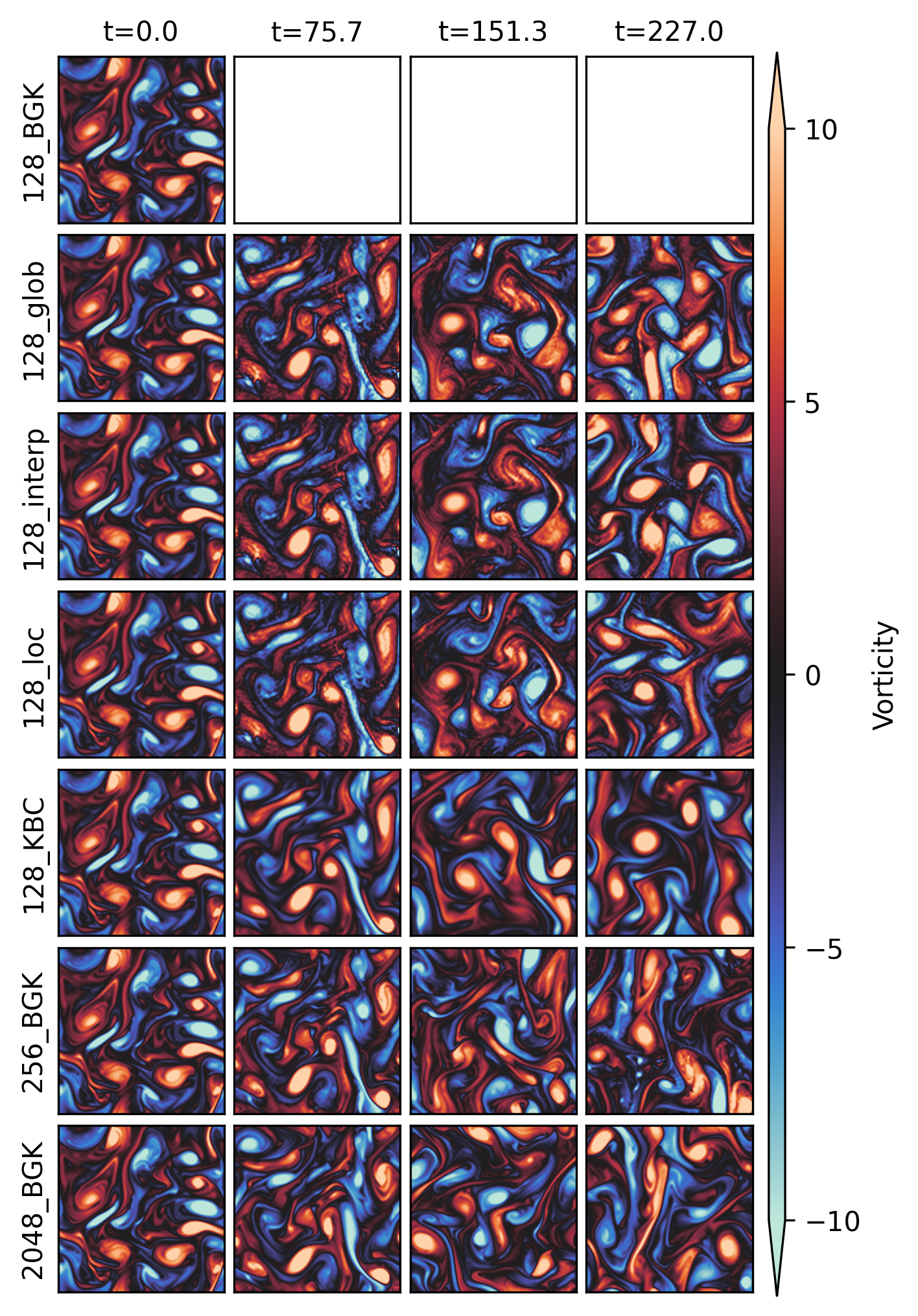}
        \caption{}
        \label{img:vorts_1e4}
    \end{subfigure}
    \begin{subfigure}[t]{0.32\textwidth}
        \centering
        \includegraphics[width=1.0\textwidth]{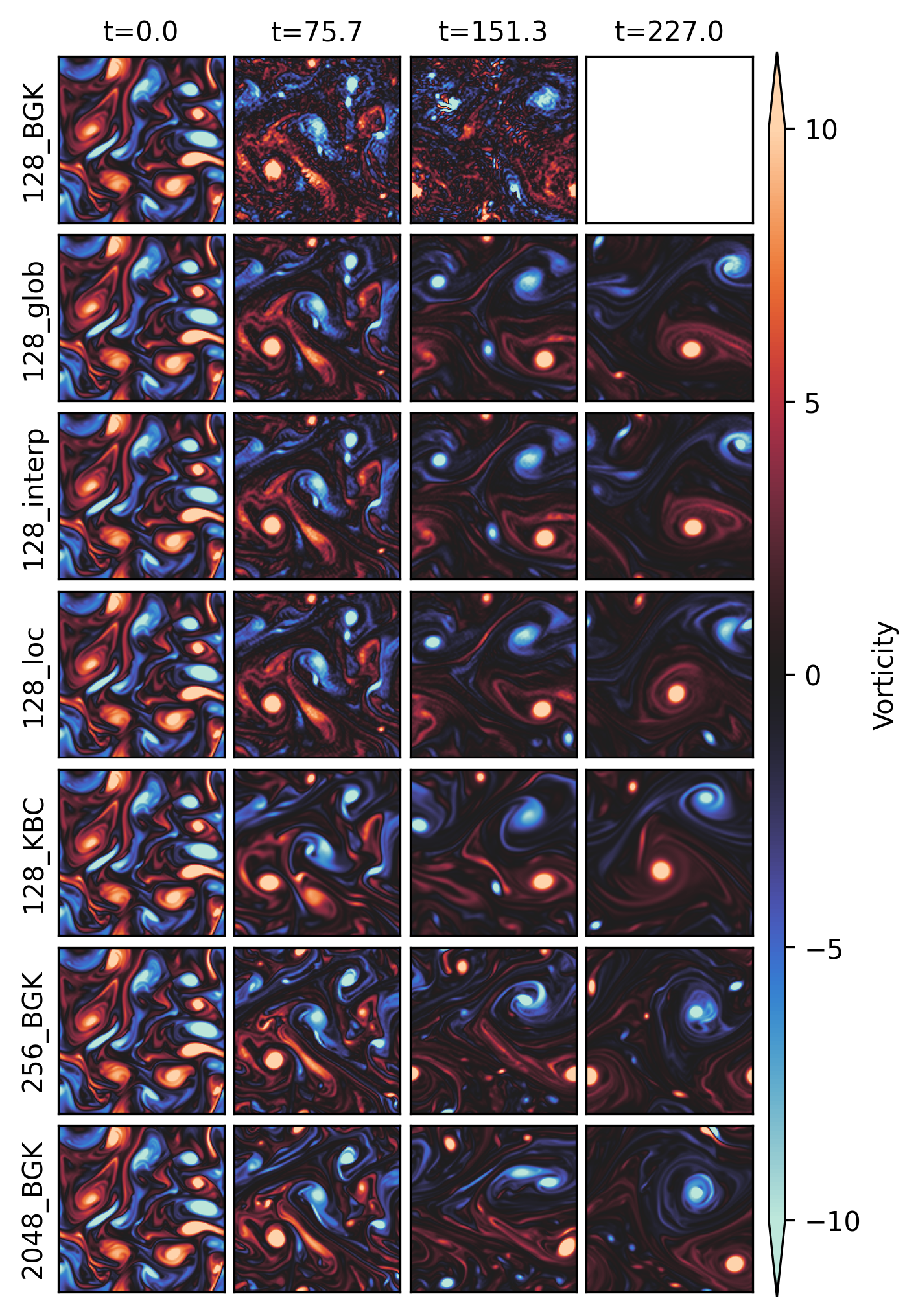}
        \caption{}
        \label{img:vorts_1e4_decay}
    \end{subfigure}
    \begin{subfigure}[t]{0.32\textwidth}
        \includegraphics[width=1.0\textwidth]{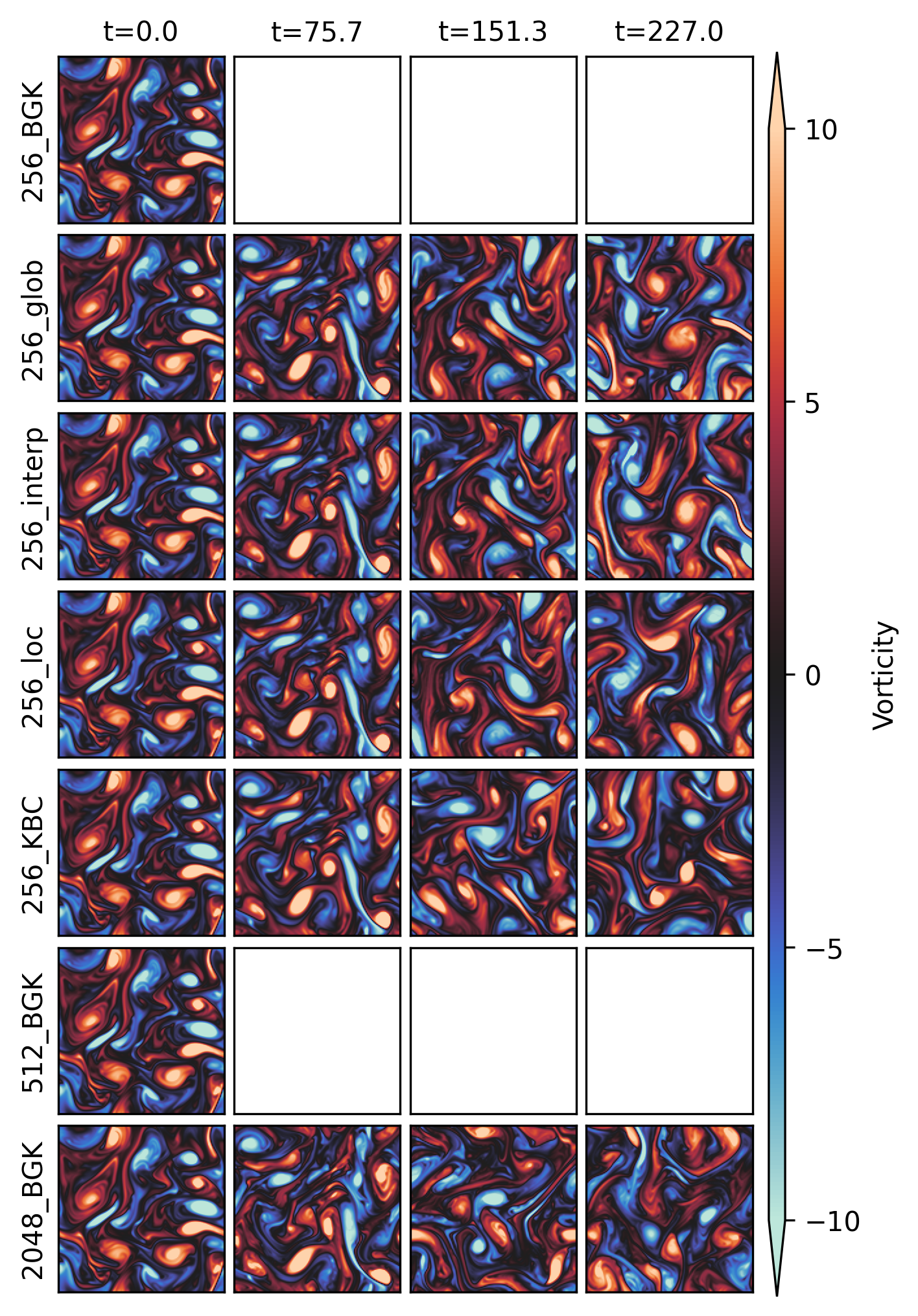}
        \caption{}
        \label{img:vorts_1e5}
    \end{subfigure}
    \caption{Comparison of the vorticity fields of all models on the three test cases: Kolmogorov flow at $Re=10^4$ (\ref{img:vorts_1e4}), decaying unforced flow at $Re=10^4$ (\ref{img:vorts_1e4_decay}), and Kolmogorov flow at $Re=10^5$ (\ref{img:vorts_1e5}).}
    \label{img:all_vorts}
\end{figure}

Figure \ref{img:1e4_results} presents the results from the first test case, where the trained models are evaluated over longer simulation times. The correlation plot in Figure \ref{img:corrs_1e4} demonstrates that all three models (global, local, and interpolated) successfully stabilize the CGS at a resolution of $128^2$, exhibiting stabilizing effects comparable to the KBC model. These models maintain stability for durations significantly longer than those encountered during training. The energy spectrum comparison, shown in Figure \ref{img:spectra_1e4}, reveals that the energy spectra of all models closely match the desired DNS spectra, with only minor deviations at high wave numbers. These deviations are small in comparison to the error seen with the KBC model at large wave numbers. Figure \ref{img:1e4_decay_results} showcases the models' performance on an unseen decaying unforced turbulent flow. All three models accurately replicate the target energy spectrum, exhibiting smaller deviations at high wave numbers compared to the KBC model. This suggests that the trained models are not only able to stabilize the flow but also generalize well to different flow conditions, producing more accurate results in terms of energy distribution. When evaluated on a Kolmogorov flow at Reynolds number $Re=10^5$ (see Fig. \ref{img:1e5_results}), the trained models proved capable of stabilizing this more turbulent flow, where even the BGK simulation at a resolution of $512^2$ becomes unstable. In line with the previous test case results, all models continued to reproduce the energy spectrum with minimal deviations at high wave numbers. This indicates that the models are robust and can handle higher Reynolds numbers, as well as more complex turbulent flows, without sacrificing stability or accuracy. Additionally, the performance on the higher resolution Kolmogorov flow ($Re=10^5$) further demonstrated the flexibility and adaptability of the trained models. The fully convolutional network architecture allowed the models to seamlessly adjust to the new grid resolution without any need for retraining or modification. Finally, Figure \ref{img:all_vorts} illustrates the evolution of the vorticity field $\omega = \partial_x u_y - \partial_y u_x$ across all models and test cases. The vorticity fields show that the trained models accurately capture the fine-scale structures of turbulence, preserving flow features over time. In summary, the results demonstrate that the trained models can successfully stabilize under-resolved Lattice Boltzmann simulations across multiple test cases, including both stationary and decaying turbulent flows, as well as high Reynolds number conditions. The models consistently reproduce the energy spectra of DNS simulations with minimal deviations, outperforming the traditional KBC and BGK models in terms of stability and accuracy, especially at high wave numbers. The differences between local, global, and interpolating agents are minimal, which strongly suggest that for most applications a purely global model is the best choice due to the lowest training time. A detailed analysis of the learned policy can be found in Appendix \ref{app:policy}.

\section{Discussion and Conclusions}

This study demonstrates the potential of RL-driven closure models for stabilizing under-resolved Lattice Boltzmann simulations. By leveraging the power of neural networks in combination with the local and physical conservation properties of the Lattice Boltzmann Method (LBM), we have developed models that  accurately capture turbulence statistics at a fraction of the cots of scale resolving simulations. The proposed closures exhibit stability for extended simulation times and are  able to generalize well across different flow typesand Reynolds numbers. A key advantage of RL in closure modeling is its posterior learning nature, allowing the models to adaptively correct numerical compounding errors without requiring expensive direct numerical simulations (DNS) during training. Leveraging the LBM's inherent exact conservation properties, locality, and straightforward access to macroscopic quantities via moments presents unique opportunities for improved hybrid closure modeling. Our results show that trained RL models significantly outperform both the KBC and BGK models, yielding a more accurate energy spectrum, particularly in the challenging high-wave number regime. Although the KBC model also stabilizes the CGS, it does not achieve this level of spectral accuracy. This is significant as accurate representation of high wave numbers is crucial for capturing fine-scale turbulent features, which can strongly influence the evolution of turbulent flows. Additionally, the models stabilize a high Reynolds number Kolmogorov flow, further showcasing the robustness and adaptability of the  approach. The versatility of the fully convolution architecture was shown by the ease of scaling to different grid resolutions without any modifications. Several extensions can further improve the proposed RL-LB methodology. Incorporating higher-order statistics as objectives can comprehensively enhance model evaluation and performance. Advanced closure models could learn more general closure formulations that do not rely on the BGK approximation, such as a closure for the collision operator or modeling the equilibrium distribution directly. Alternatively, employing a local reward mechanism could better exploit the locality, addressing the credit assignment problem more effectively. 
The ongoing work aims to extend the RL-LBM to flows with solid boundaries and to flows in porous media.  

In summary, we have shown that reinforcement learning effectively discovers closure models that stabilize under-resolved Lattice-Boltzmann simulations. By leveraging LBM, especially in its entropic version, and RL, our models stabilize CGS simulations, generalize to new flow cases, and outperform traditional models like KBC and BGK, especially in high-wave number energy spectrum areas. The flexibility of our RL framework and scalable neural architecture make it suitable for large-scale turbulent simulations across various grid resolutions. RL-driven closure models can enhance the efficiency and accuracy of such simulations, particularly when DNS data are costly or unavailable. 
This work advances the integration of RL models in fluid dynamics, paving the way to high-fidelity simulations using the Lattice-Boltzmann method for aerodynamics,  flows in porous media, and microfluidics.


\bibliography{bibliography}
\clearpage
\appendix
\input{appendix}

\end{document}

%% file: Appendix.tex

\section{Kolmogorov Flow implementation details}\label{sec:kolmgorov-appendix}

The Kolmogorov flow is a flow described by the incompressible Navier-Stokes equations in 2d on the domain $D=[0,l]\times[0,l]$ with periodic boundary conditions and Kolmogorov forcing $f_x = \chi sin(\kappa y)$, where $\kappa = \frac{2\pi n}{l}$ \citep{Kolmogorov_flow_definition}. An artifact of forced flows in 2d is an inverse energy cascade. This can be counteracted by introducing a friction term such that the total force has the form $\boldsymbol{f} = \chi sin(\kappa y)\hat{\boldsymbol{x}} - \alpha \boldsymbol{u}$, where $\hat{\boldsymbol{x}}$ is the unit vector in x-direction \citep{2d-turbulence}. For convenience and to follow the notation of \citep{Brenner}, we will choose the following parameters: 
$$ l = 2\pi,\quad \chi = 1,\quad \alpha = 0.1.$$
We note that these parameters are choosen to be dimensionless and can be transformed via $dx=\frac{l}{2\pi}$ and $dt=\sqrt{\frac{l}{2\pi\chi}}$.

Taking $$ Re = \frac{UL}{\nu} \equiv \frac{\sqrt{\chi}}{\nu}\left( \frac{l}{2\pi} \right)^{3/2}$$ from \citep{Kolmogorov_flow_definition} we get:
$$ \nu = \frac{1}{Re}, \quad U = n, \quad L = \frac{1}{n}.$$
Lattice Boltzmann simulations are typically performed in lattice units (indicated by an asterisk) for which a complete set of conversion factors are chosen $\{C_l = \Delta x, C_t=\Delta t, C_\rho = \rho\}$, s.t. the following relations hold in lattice units $\Delta t^* = \Delta x^* = \rho^* \equiv 1$ and $c_s^2 = \frac{1}{3}$. Conversion are performed via $l^* = \frac{l}{C_l}$.
To transform to lattice units we use
$$\Delta x = \frac{l}{N} = \frac{2\pi}{N},\quad \Delta t = \frac{\nu^*}{\nu}\Delta x^2.$$

From the law of similarity, stating that the Reynolds number computed in physical and in lattice units must match, we get $$\nu^* =\nu U^* L^*.$$ Here $L^* = \frac{N}{2\pi n}$ and $U^*$ can be chosen within stability and accuracy bounds \citep{LBM_Krueger}, e.g. low mach number hypothesis. To satisfy this we will choose $U^* = 0.1c_s^*$. With this we can transform all parameters and obtain: 
$$\chi^* = \frac{2\pi\chi}{N}\left(\frac{U^*}{n}\right)^2, \quad \alpha^* =\frac{\alpha n}{\chi U^*}\chi^*.$$

Following \citep{Brenner}, a time-step $\delta t = \frac{1}{2}\frac{\Delta x}{v_{max}}$ with $v_{max} = 7$ is used. Taking $m = \frac{t}{\delta t}$ steps corresponds to $m^* = \frac{m}{\tau}$ LBM steps, with $\tau = 2 v_{max}\frac{U^*}{n}$. This corresponds to the non-dimensional time $T = m^*\frac{U^*}{L^*}$. We note that we initialize the velocity field randomly as described in \citep{initialization_kolmogorov} and use $v_{max}^* = \frac{U^*}{n}v_{max}$. 

\section{Energy Spectrum with Standard Deviation}\label{sec:std_spectra}

\begin{figure}[H]
    \centering
    \begin{subfigure}[t]{0.48\textwidth}
        \centering
        \includegraphics[width=\textwidth]{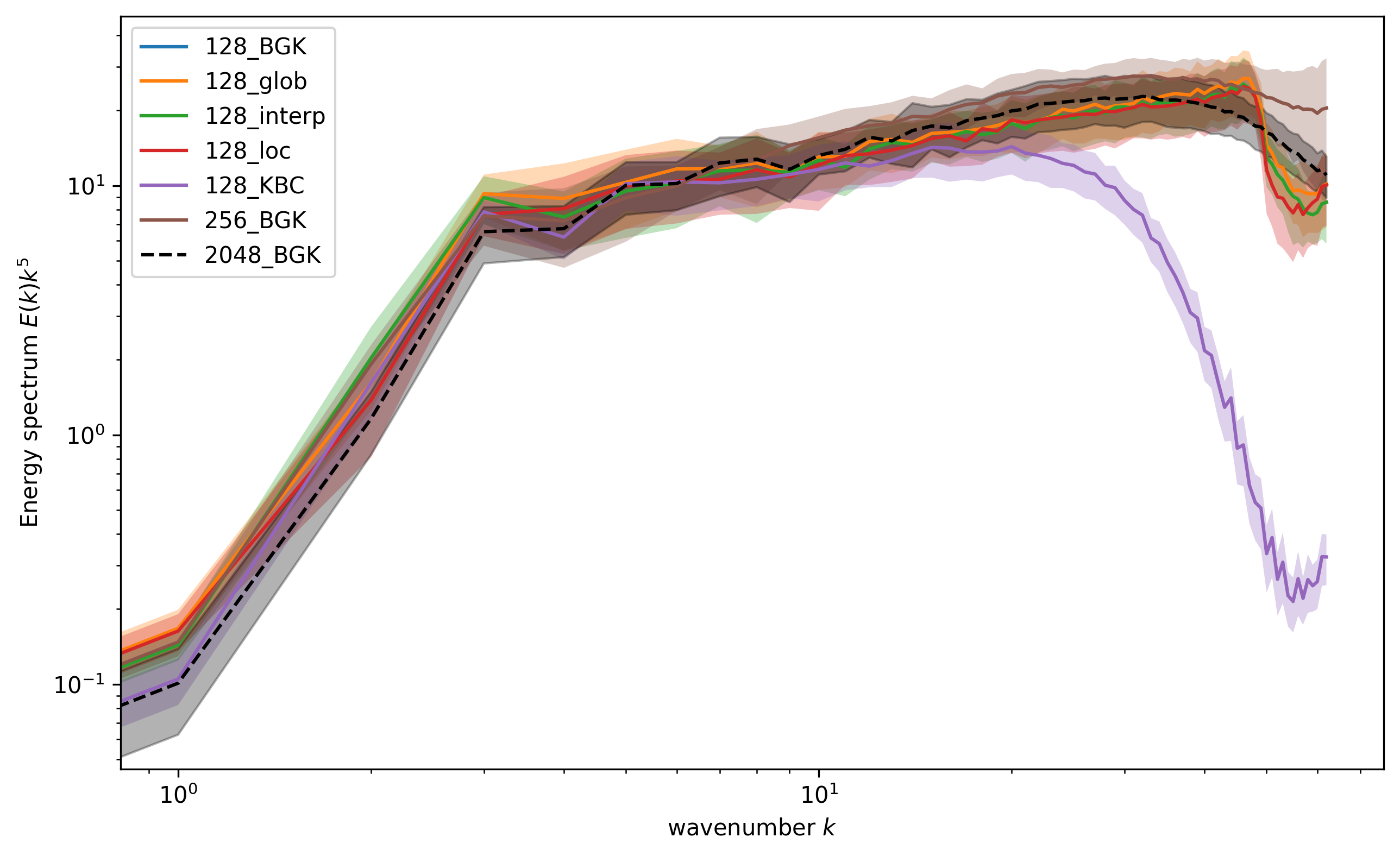}
        \caption{}
        \label{img:std_spectra_1e4}
    \end{subfigure}
    \begin{subfigure}[t]{0.48\textwidth}
        \centering
        \includegraphics[width=\textwidth]{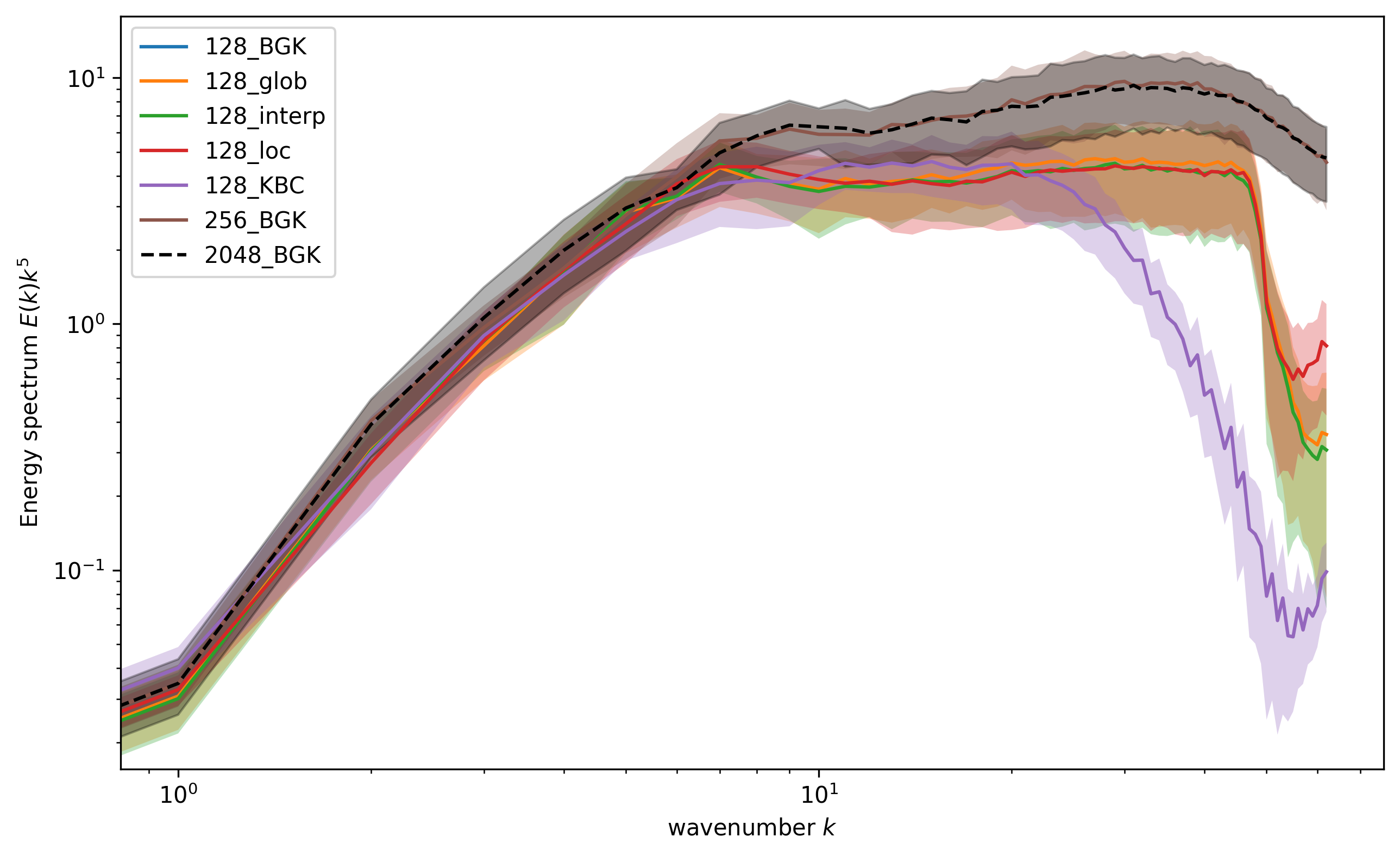}
        \caption{}
        \label{img:std_spectra_1e4_decay}
    \end{subfigure}
    \begin{subfigure}[t]{0.48\textwidth}
        \centering
        \includegraphics[width=\textwidth]{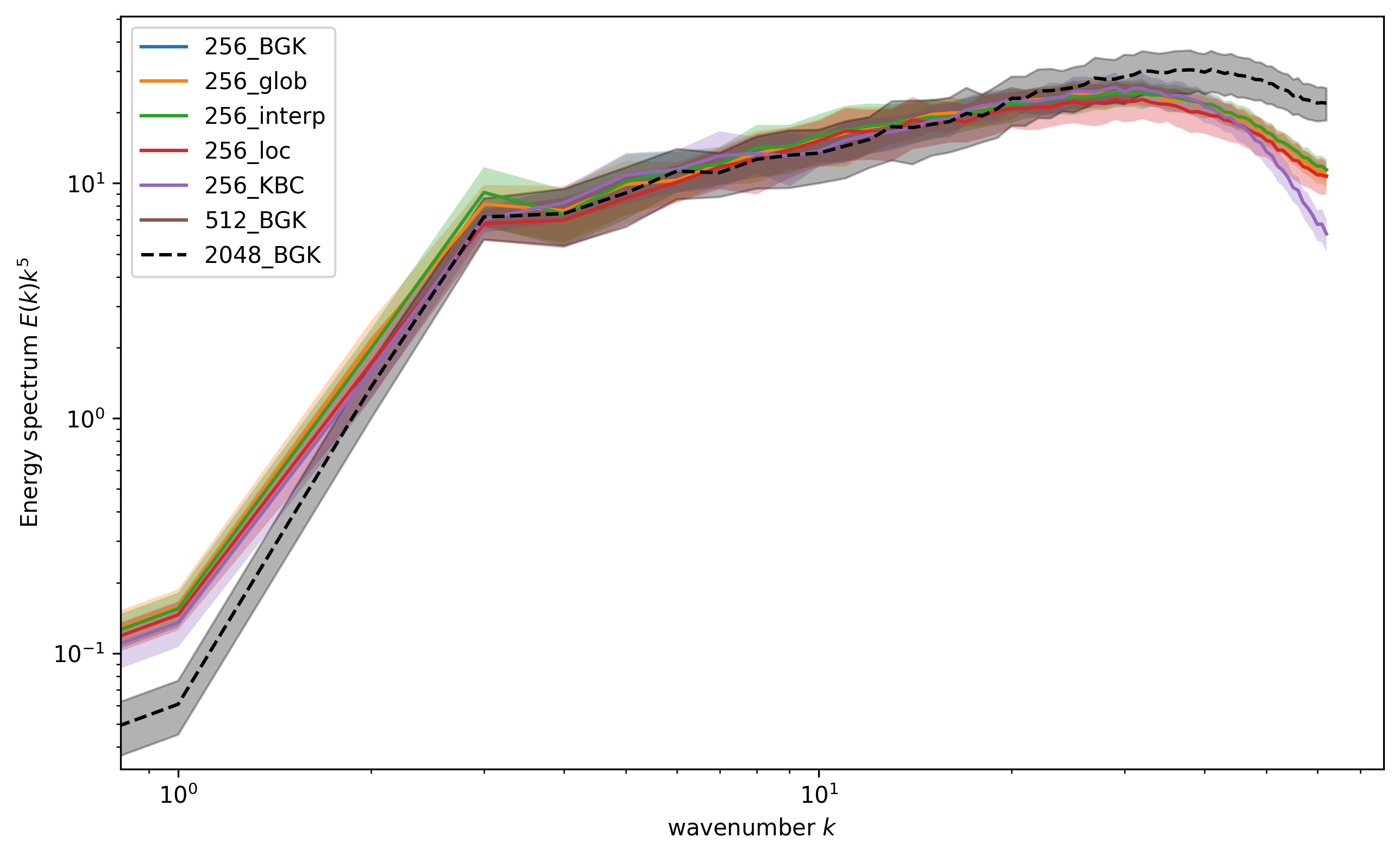}
        \caption{}
        \label{img:std_spectra_1e5}
    \end{subfigure}
    \caption{Energy spectra plots scaled by $k^5$, showing the mean and standard deviation computed over the second half of the simulation $T\in[113, 227]$. The subplots show the spectra for the three test cases: Kolmogorov flow at $Re=10^4$ (\ref{img:std_spectra_1e4}), decaying unforced flow at $Re=10^4$ (\ref{img:std_spectra_1e4_decay}), and Kolmogorov flow at $Re=10^5$ (\ref{img:std_spectra_1e5}).}
    \label{img:std_spectra}
\end{figure}

\section{Training Details and Hyperparameters}\label{sec:App_Hyp}
The neural network architectures for global, local, and interpolating policy networks are shown in tables \ref{tab:glob_actor}, \ref{tab:loc_actor}, \ref{tab:interp_actor}. The neural network architecture for the critic network is shown in Table \ref{tab:critic}, and the hyperparameters used to train these networks are shown in Table \ref{tab:hyperparameters}. We trained all networks on one NVIDIA A100 80GB GPU, where training took approximately one hour for global, three hours for local, and five hours for interpolating agents. The increase in training time for the interpolating agents is due to the overhead of interpolating actions. 

\begin{table}[ht]
\centering
\begin{tabular}{|c|c|c|c|}
    \toprule
    Hyperparameters & global & local & interpolating \\
    \midrule
    seed & 66 & 44 & 33 \\
    step\_factor & 8 & 8 & 4 \\
    num\_agents & 1 & 128 & 16 \\
    learning\_rate & 0.001 & 0.001 & 0.001 \\
    adam\_eps & $10^{-7}$ & $10^{-7}$ & $10^{-7}$  \\
    gamma & 0.99 & 0.99 & 0.99 \\
    reward\_normalization & True & True & True \\
    advantage\_normalization & True & True & True \\
    recompute\_advantage & False & False & False \\
    deterministic\_eval & True & True & True \\
    value\_clip & True & True & True \\
    action\_scaling & True & True & True \\
    action\_bound\_method & clip & clip & clip \\
    ent\_coef & -0.01 & 0 & 0 \\
    vf\_coef & 0.25 & 0.25 & 0.25 \\
    clip\_range & 0.2 & 0.2 & 0.2 \\
    max\_grad\_norm & 0.5 & 0.5 & 0.5 \\
    gae\_lambda & 0.95 & 0.95 & 0.95 \\
    lr\_decay & False & True & False \\
    buffer\_size & 2000 & 2000 & 2000 \\
    max\_epoch & 100 & 300 & 200 \\
    step\_per\_epoch & 1500 & 1500 & 1500 \\
    repeat\_per\_collect & 3 & 3 & 3 \\
    episode\_per\_test & 1 & 1 & 1 \\
    batch\_size & 64 & 64 & 64 \\
    step\_per\_collect & 128 & 128 & 128 \\
    \bottomrule
\end{tabular}
\caption{Hyperparameters used for training global local and interpolating models. For description see Tianshou \cite{tianshou} documentation.}
\label{tab:hyperparameters}
\end{table}

\begin{table}[ht]
\centering
\begin{tabular}{|l|l|}
\toprule
Layer Name & Details \\
\midrule
model.0 & Conv2d(6, 128, kernel\_size=(1, 1), stride=(1, 1)) \\
model.1 & ReLU(inplace=True) \\
model.2 & Conv2d(128, 128, kernel\_size=(1, 1), stride=(1, 1)) \\
model.3 & ReLU(inplace=True) \\
mu.0 & Conv2d(128, 1, kernel\_size=(1, 1), stride=(1, 1)) \\
mu.1 & Tanh() \\
sigma.0 & Conv2d(128, 1, kernel\_size=(1, 1), stride=(1, 1)) \\
sigma.1 & Softplus(beta=1.0, threshold=20.0) \\
\bottomrule
\end{tabular}
\caption{Neural network architecture of vectorized local policy network.}
\label{tab:loc_actor}
\end{table}

\begin{table}[ht]
\centering
\begin{tabular}{|l|p{10cm}|}
\toprule
Layer Name & Details \\
\midrule
model.0 & Conv2d(6, 64, kernel\_size=(9, 9), stride=(4, 4), padding=(4, 4), padding\_mode=circular) \\
model.1 & ReLU(inplace=True) \\
model.2 & Conv2d(64, 64, kernel\_size=(5, 5), stride=(2, 2), padding=(2, 2), padding\_mode=circular) \\
model.3 & ReLU(inplace=True) \\
model.4 & Conv2d(64, 64, kernel\_size=(3, 3), stride=(2, 2), padding=(1, 1), padding\_mode=circular) \\
model.5 & ReLU(inplace=True) \\
model.6 & Conv2d(64, 64, kernel\_size=(3, 3), stride=(1, 1), padding=(1, 1), padding\_mode=circular) \\
model.7 & ReLU(inplace=True) \\
model.8 & MaxPool2d(kernel\_size=2, stride=2, padding=0,) \\
model.9 & Conv2d(64, 64, kernel\_size=(3, 3), stride=(1, 1), padding=(1, 1), padding\_mode=circular) \\
model.10 & ReLU(inplace=True) \\
model.11 & MaxPool2d(kernel\_size=2, stride=2,) \\
fcnn.0 & Linear(in\_features=256, out\_features=128, bias=True) \\
fcnn.1 & ReLU(inplace=True) \\
fcnn.2 & Linear(in\_features=128, out\_features=64, bias=True) \\
fcnn.3 & ReLU(inplace=True) \\
mu.0 & Linear(in\_features=64, out\_features=1, bias=True) \\
mu.1 & Tanh() \\
sigma.0 & Linear(in\_features=64, out\_features=1, bias=True) \\
sigma.1 & Softplus(beta=1.0, threshold=20.0) \\
\bottomrule
\end{tabular}
\caption{Neural network architecture of global policy network.}
\label{tab:glob_actor}
\end{table}

\begin{table}[ht]
\centering
\begin{tabular}{|l|p{10cm}|}
\toprule
Layer Name & Details \\
\midrule
model.0 & Conv2d(6, 128, kernel\_size=(9, 9), stride=(8, 8), padding=(1, 1), padding\_mode=circular) \\
model.1 & ReLU() \\
model.2 & Conv2d(128, 128, kernel\_size=(1, 1), stride=(1, 1)) \\
model.3 & ReLU() \\
model.4 & Conv2d(128, 128, kernel\_size=(1, 1), stride=(1, 1)) \\
model.5 & ReLU() \\
mu.0 & Conv2d(128, 1, kernel\_size=(1, 1), stride=(1, 1)) \\
mu.1 & Tanh() \\
sigma.0 & Conv2d(128, 1, kernel\_size=(1, 1), stride=(1, 1)) \\
sigma.1 & Softplus(beta=1.0, threshold=20.0) \\
\bottomrule
\end{tabular}
\caption{Neural network architecture of vectorized interpolating policy network.}
\label{tab:interp_actor}
\end{table}

\begin{table}[ht]
\centering
\begin{tabular}{|l|p{10cm}|}
\toprule
Layer Name & Details \\
\midrule
model.0 & Conv2d(6, 64, kernel\_size=(9, 9), stride=(4, 4), padding=(4, 4), padding\_mode=circular) \\
model.1 & ReLU(inplace=True) \\
model.2 & Conv2d(64, 64, kernel\_size=(5, 5), stride=(2, 2), padding=(2, 2), padding\_mode=circular) \\
model.3 & ReLU(inplace=True) \\
model.4 & Conv2d(64, 64, kernel\_size=(3, 3), stride=(2, 2), padding=(1, 1), padding\_mode=circular) \\
model.5 & ReLU(inplace=True) \\
model.6 & Conv2d(64, 64, kernel\_size=(3, 3), stride=(1, 1), padding=(1, 1), padding\_mode=circular) \\
model.7 & ReLU(inplace=True) \\
model.8 & MaxPool2d(kernel\_size=2, stride=2) \\
model.9 & Conv2d(64, 64, kernel\_size=(3, 3), stride=(1, 1), padding=(1, 1), padding\_mode=circular) \\
model.10 & ReLU(inplace=True) \\
model.11 & MaxPool2d(kernel\_size=2, stride=2) \\
fcnn.0 & Linear(in\_features=256, out\_features=128, bias=True) \\
fcnn.1 & ReLU(inplace=True) \\
fcnn.2 & Linear(in\_features=128, out\_features=64, bias=True) \\
fcnn.3 & ReLU(inplace=True) \\
fcnn.4 & Linear(in\_features=64, out\_features=1, bias=True) \\
\bottomrule
\end{tabular}
\caption{Neural network architecture of central critic network.}
\label{tab:critic}
\end{table}

\newpage
\section{Analysis of the learned policy}
\label{app:policy}
\begin{figure}[h]
    \centering
    \includegraphics[width=0.7\textwidth]{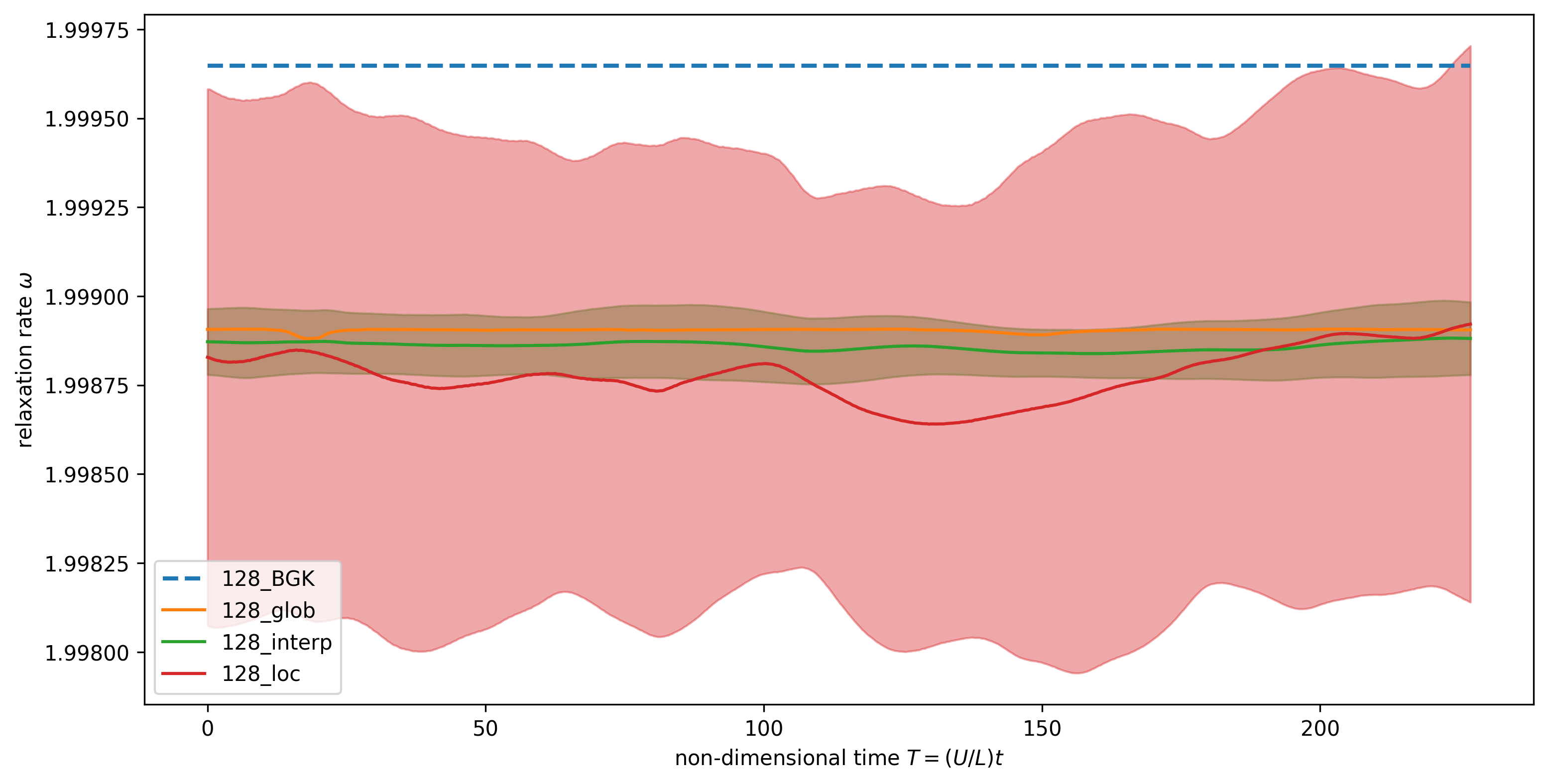}
    \caption{Mean and standard deviation (over the whole domain) of relaxation rates computed from the actions of the local global and interpolating models, for a Kolmogorov flow at $Re=10^4$ for $T=227$. 
    No standard deviations are available for the BGK as well as the global RL model.}
    \label{img:relaxation-rate}
\end{figure}

The computed relaxation rates for the global, local, and interpolating models can be seen in Figure \ref{img:relaxation-rate}. The mean and standard deviation are computed over space and plotted over time for a Kolmogorov flow at $Re=10^4$ and $T=227$. The global model attenuates the BGK relaxation rate to an approximately constant value. Since the global model predicts a global action, no standard deviation is shown. The interpolating and fully local models produce a similar mean relaxation rate to the global model. The local model shows more variation in the mean and a larger standard deviation than the interpolating model. The change of the relaxation rate $w$ is expected \citep{succi2018lattice} and a smaller relaxation rate leads to increased stability of the simulation. Other approaches \citep{chen2003extended} have also suggested closure models for LB that modify the relaxation rate.

It is worth emphasizing that the correction to the nominal value of the relaxation frequency may appear numerically negligible (fourth digit), but it is definitely not, the reason being that the simulation 
operates in the vicinity of the zero viscosity limit, where small changes in $\omega$ result in large
changes of $2-\omega$. For instance taking $\omega_0=1.99965$ as a nominal BGK value and 
$\omega_{g}=1.99891$ as a representative value after reinforcement learning with global agents, leads to 
an effective viscosity $\nu_0 \sim 1.45 \cdot 10^{-6}$ versus $\nu_g \sim 4.5 \cdot 10^{-6}$, respectively. 
This is a pretty sizeable factor three increase, which accounts for the observed stabilization effect.

It is also worth noting that despite the significant increase of the effective viscosity as compared to the nominal
value, the simulations remain in the strongly over-relaxed regime $(2-\omega) \ll 1$ corresponding
to near-equilibrium turbulence characterised by a neat separation of scales between large and small eddies, so
that the notion of an effective eddy-viscosity still retains a well-defined physical meaning.     
Future work will be devoted to explore whether the reinforcement learning strategy discussed in this paper
can also be applied successfully to the case of strong non-equilibrium turbulence, such as the one that occurs
for high Reynolds flows confined by solid boundaries.

\begin{figure}[H]
    \centering
    \begin{subfigure}[t]{0.49\textwidth}
    \includegraphics[width=1.0\textwidth]{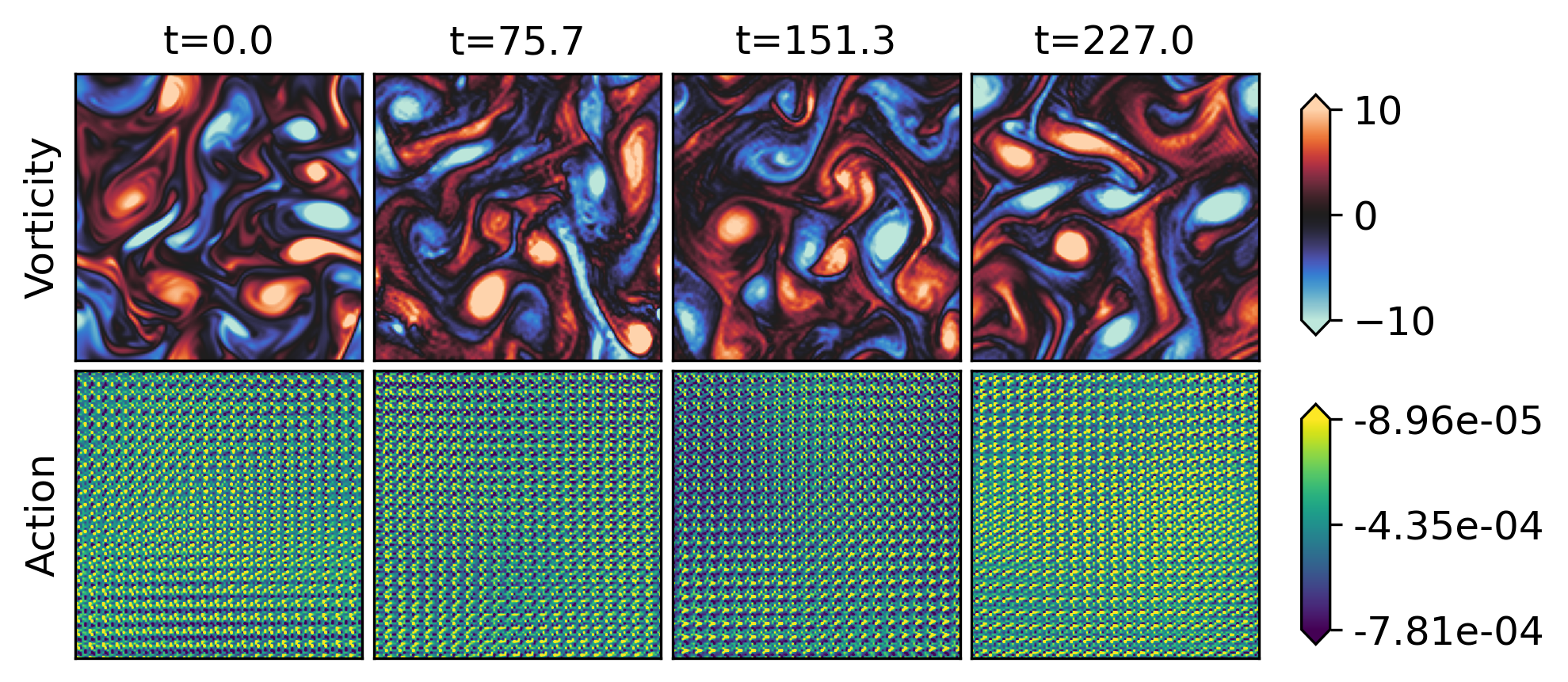}
        \caption{}
        \label{img:actions-loc}
    \end{subfigure}
    \begin{subfigure}[t]{0.49\textwidth}
        \centering
        \includegraphics[width=1.0\textwidth]{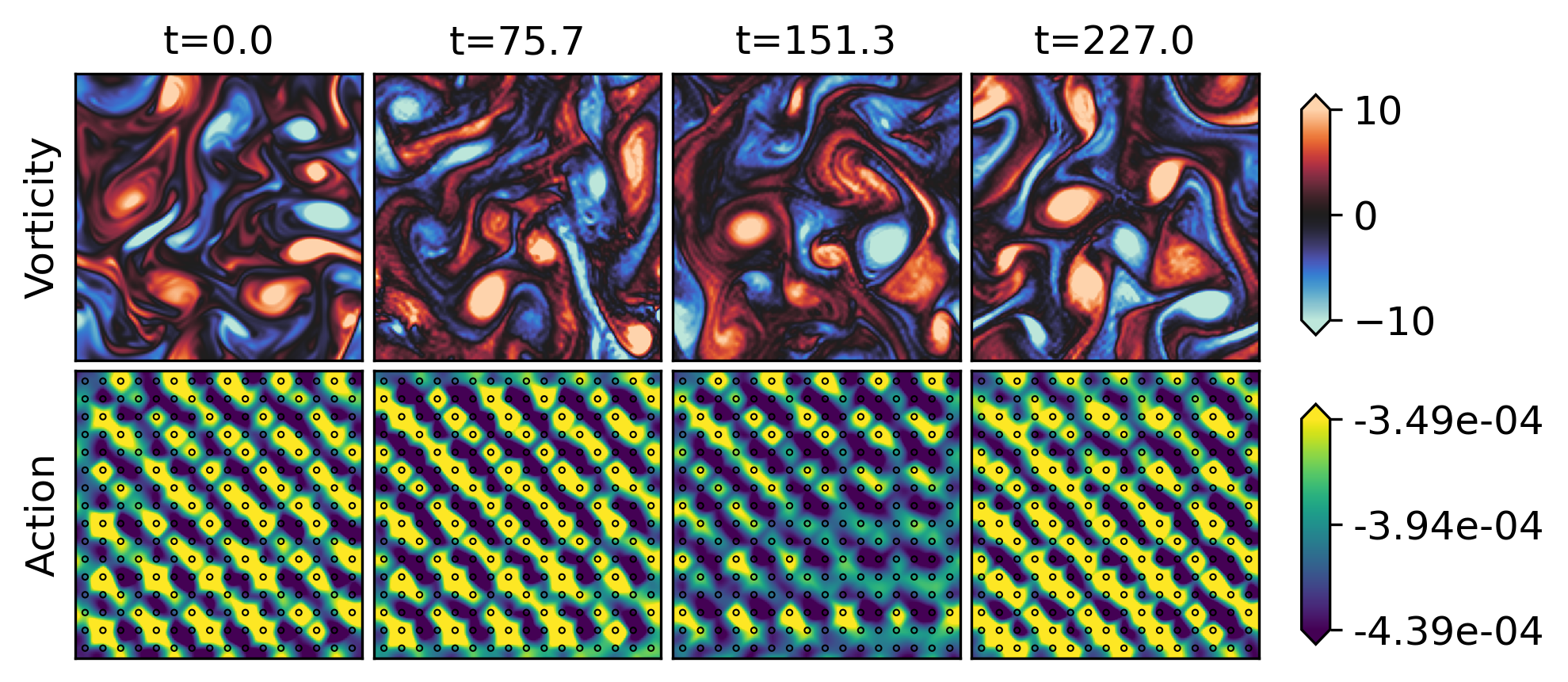}
        \caption{}
        \label{img:actions-interp}
    \end{subfigure}
    \caption{Action visualization of the local \ref{img:actions-loc} and interpolating \ref{img:actions-interp} models. Four plots showing the actions taken at each grid point and the corresponding vorticity fields at four equidistant times during the simulation of a Kolmogorov flow at $Re=10^4$.}
    \label{img:actions-over-time}
\end{figure}

Visual examples of actions for the local and interpolating models are shown in Figure \ref{img:actions-over-time}, where the actions at each grid point are plotted at four different times during the simulation of a Kolmogorov flow at $Re=10^4$. The action values shown are the network outputs, which are in $[-\epsilon, \epsilon]$. To calculate the relaxation rate, they have been transformed to $[2-\epsilon, 2+\epsilon]$.
